\documentclass[11pt]{article}
\begin{document}
\title{Networks of self-avoiding chains
and Ogden-type constitutive equations for elastomers}

\author{A.D. Drozdov and M. Gottlieb\\
Department of Chemical Engineering\\
Ben-Gurion University of the Negev\\
P.O. Box 653\\
Beer-Sheva 84105, Israel}
\date{}
\maketitle

\begin{abstract}
An expression is derived for the strain energy of
a polymer chain under an arbitrary three-dimensional
deformation with finite strains.
For a Gaussian chain, this expression is reduced
to the conventional Moony--Rivlin constitutive law,
while for non-Gaussian chains it implies
novel constitutive relations.
Based on the three-chain approximation,
explicit formulas are developed for the
strain energy of a chain modeled as a self-avoiding
random walk.
In the case of self-avoiding chains
with stretched-exponential distribution function
of end-to-end vectors, the strain energy
density of a network is described by
the Ogden law with only two material constants.
For the des Cloizeaux distribution function,
the constitutive equation involves three 
adjustable parameters.
The governing equations are verified by fitting
observations on uniaxial tension, uniaxial
compression and biaxial tension of elastomers.
Good agreement is demonstrated between the
experimental data and the results of numerical
analysis.
An analytical formula is derived for the ratio
of the Young's modulus of a self-avoiding chain
to that of a Gaussian chain.
It is found that the elastic modulus per chain
in the Ogden network exceeds that in a Gaussian
network by a factor of three, whereas the
elastic modulus of a chain with the generalized
stretched exponential distribution function equals
about half of the modulus of a Gaussian chain.
\end{abstract}
\vspace*{10 mm}
\noindent
{\bf Key-words:}
Constitutive equations,
Elastomers,
Affine networks,
Non-Gaussian chains,
Self-avoiding random walks
\newpage

\section{Introduction}

This paper is concerned with the so-called
``physically motivated" constitutive equations
for the elastic response of elastomers
at finite strains.
Although this subject has attracted substantial
attention in the past half a century,
it remains a focus of attention
both in the communities of mechanical
engineers,
\cite{BB98,KS00,CCH01,AAO02,BAG02,EZB02,Bea03,CC03,HTS03,MM03}
and polymer physicists
\cite{BM98,SB98,BWA99,BM00,GL00,SGB01,CMM02,SS02,SN03,Win03,LT04}.
This interest may be explained by two reasons:
(i) from the standpoint of applications,
it is induced by the necessity to have a reliable
tool for the analysis of the mechanical
behavior of polymer materials,
whose response is inadequately described by
conventional stress--strain equations,
and (ii) from the point of view of fundamental
research, it is driven by the lack of
constitutive laws based on the solid ground
of statistical physics that provide rather simple
formulas for the strain energy of rubber-like materials.

Surprisingly, the concept of Gaussian chains
in an affine network introduced about a century
ago \cite{Tre75} may serve as the only example
of a physically-based theory that can be employed
in the analysis of stresses in
elastomer structures (in this assertion,
we exclude from the consideration
(i) slip-link models \cite{EV88} that deal with
chains in non-affine networks,
(ii) numerous variants, see \cite{BA00} the the references
therein, of the
Kratky--Porod\footnote{We employ the terminology
widely used in statistical mechanics of macromolecules.
In the studies on mechanics of polymers, this model is
conventionally referred to as the James--Guth theory\cite{JG43}
or the theory of Langevin chains \cite{MM03}.}
model \cite{KP49}, where the stress--strain
relations contain the inverse Langevin function
and can be resolved only numerically,
and (iii) the Gent \cite{Gen96}, see also \cite{HS02},
and FENE dumbbell \cite{HO97} constitutive equations,
which correctly mimic a limiting chain
extensibility, but can be deduced from an appropriate
statistical theory within the Peterlin approximation only).
The model of a Gaussian chain is, however, overly
simplified for engineering applications, because it
implies the neo--Hookean expression
for the strain energy per chain
\begin{equation}
W=\frac{1}{2}k_{\rm B}T (J_{1}-3),
\end{equation}
which rather poorly describes observations.
Here $k_{\rm B}$ is Boltzmann's constant,
$T$ is the absolute temperature,
and $J_{m}$ stands for the $m$th principal
invariant of the right Cauchy--Green
deformation tensor ${\bf C}$.
From the physical standpoint, other shortcomings
of the model of a freely jointed chain are
that it
(i) disregards short- and long-range interactions
between segments,
and (ii) implies that the end-to-end distance
exceeds the contour length of a chain with a
non-zero probability.

To avoid these disadvantages, more sophisticated
models of non-Gaussian linear chains are introduced.
The starting point in their derivation
is the treatment of a chain as a random curve
in a three-dimensional space \cite{DE86}.
The randomness of the curve reflects random
fluctuations of the chain driven by thermal excitations.
Each realization of the curve is described by the
vector equation $\tilde{{\bf Q}}(s)$, where $s$
is the arc-length.
For definiteness, we suppose that $s\in [0,L]$,
where $L$ stands for the contour length,
and assume the end $s=0$ to be fixed at the origin,
and the end $s=L$ to be free.
In these notation, the end-to-end vector of a chain
reads ${\bf Q}=\tilde{\bf Q}(L)$.

As the description of random curves is extremely
complicated from the mathematical standpoint,
it is convenient to replace any curve by a set
of $N$ rigid segments with length $l$
connected with each other.
These segments are thought of as random vectors,
which are independent for Gaussian chains or
mutually dependent for more advanced models.
The segment length $l$ and the number of segments
$N$ are connected by the formula
$L=Nl$.
Conventionally, $N$ is treated as a given integer
number.
The parameter $l$ may be fixed (for the Gaussian
and Kratky--Porod chains),
or it may be considered as a random variable
(for the L$\acute{\rm e}$vy flights \cite{Lev54}).
For a fixed $l$, the discrete model of a chain is
equivalent to a random walk with $N$ steps
that starts at the origin.

Formally, a random walk is entirely described by
the probability density
$f(\tilde{\bf Q}_{1},\ldots,\tilde{\bf Q}_{N})$,
where the vector $\tilde{\bf Q}_{n}$ characterizes
the $n$th step of the walk.
For a large $N$, this description becomes, however,
inconvenient,
and only the distribution $p_{0}({\bf Q})$
of the corresponding end-to-end vector ${\bf Q}$
is studied.
For a spherically symmetric walk (no preferable
direction in space), the average value of
${\bf Q}$ vanishes, $\overline{\bf Q}={\bf 0}$,
whereas the mean-square end-to-end distance
$b^{2}=\overline{Q^{2}}$ characterizes the spatial
dimension of a chain.
The scaling analysis \cite{deG79} implies that
\begin{equation}
b=lN^{\nu},
\end{equation}
where $\nu$ is a scaling exponent.
For a Gaussian chain without interaction between
segments, $\nu=\frac{1}{2}$, but for a chain that
is sensitive to such interactions, $\nu$ becomes
higher.

All theories of polymer chains may be divided into
two groups:
(i) phantom chains, for which two segments of a
chain may be located at the same point,
and (ii) self-avoiding chains, for which excluded
volume interactions are taken into account.
The first group of models is relatively simple, and a
number of mathematically strong results have been
derived for phantom chains.
The most famous models of this group are
the Kratky--Porod concept \cite{KP49} that accounts for
interactions between nearest neighbors only,
and its continuous analog, the model of worm-like
chains \cite{KC82}.

Unlike most previous studies concerned with
phantom chains, the present work concentrates on
constitutive equations for linear macromolecules
modeled as self-avoiding random walks (SARW)
\cite{Van98}.
Although an exact expression for the distribution
function of end-to-end vectors of a SARW is unknown,
two convenient approximations are widely used
\cite{BC91,DP91}.
According to the first, the distribution function
of end-to-end vectors is described
by the stretched exponential expression
\begin{equation}
p_{0}({\bf Q})=p^{0}\exp \Bigl [
-\beta \Bigl (\frac{Q}{R}\Bigr )^{2\delta}\Bigr ],
\end{equation}
where $Q=|{\bf Q}|$,
and $\beta$, $\delta$, $R$ are positive constants.
The pre-factor $p^{0}$ is found from
the normalization condition
\begin{equation}
\int p_{0}({\bf Q})d{\bf Q}=1,
\end{equation}
which implies that
\begin{equation}
p^{0}=\frac{1}{4\pi} \biggl [ \int_{0}^{\infty}
\exp \Bigl (-\beta \Bigl (\frac{Q}{R}\Bigr )^{2\delta}\Bigr )
Q^{2}dQ\biggr ]^{-1}.
\end{equation}
The other approximation of the distribution
function was developed in \cite{MM71,Clo74}
\begin{equation}
p_{0}({\bf Q})=p^{0}\Bigl (\frac{Q}{R}\Bigr )^{2\alpha}
\exp \Bigl [-\beta \Bigl (\frac{Q}{R}\Bigr )^{2\delta}
\Bigr ].
\end{equation}
In the literature on random walks \cite{BC91,DP91},
Eq. (6) is conventionally referred to as the
des Cloizeaux law.
In the field of mathematical statistics,
this formula is known  \cite{LB98} as a symmetrical
Kotz-type distribution.
Equation (6) involves four adjustable constants,
$\alpha$, $\beta$, $\delta$ and $R$.
The pre-factor $p^{0}$ is determined from
the normalization condition (4),
\begin{equation}
p^{0}=\frac{R^{2\alpha}}{4\pi}
\biggl [ \int_{0}^{\infty}
\exp \Bigl (-\Bigl ( \frac{Q}{R}\Bigr )^{2\delta}
\Bigr ) Q^{2(1+\alpha)}dQ\biggr ]^{-1}.
\end{equation}
There is a substantial difference between the parameters
$\alpha$, $\beta$ and $R$, on the one hand,
and the exponent $\delta$ in Eqs. (3) and (6),
on the other.
While the quantities $\alpha$, $\beta$ and $R$
can adopt practically arbitrary values,
the parameter $\delta$ is strongly connected \cite{BC91}
with the scaling exponent $\nu$ in Eq. (2),
\begin{equation}
\delta=\frac{1}{1-\nu}.
\end{equation}
Due to some technical peculiarities of the
renormalization-group method \cite{Fre87}
employed for the evaluation of the scaling exponent,
it is convenient to assess $\nu$ for
a SARW in a $d$-dimensional space.
Flory \cite{Flo49} was the first who conjectured
that
\begin{equation}
\nu=\frac{d}{d+2}.
\end{equation}
For a three-dimensional space, Eqs. (8) and (9) imply
that $\nu=\frac{3}{5}$ and
\begin{equation}
\delta=\frac{5}{2}.
\end{equation}
Although Eq. (9) has not been either proved or disproved
up to now, the estimate (10) is used in the present
study for two reasons:
(i) formula (9) is widely employed in the field
of statistical physics of polymers,
and (ii) Eq. (10) allows necessary calculations
to be performed explicitly.

The objective of this study is three-fold:
\begin{enumerate}
\item
To derive an analytical expression for the strain
energy $W$ of a polymer chain with an arbitrary
distribution function of end-to-end vectors
$p_{0}({\bf Q})$.

\item
To apply this formula in order to develop
explicit expressions for the strain energy
of chains with distribution functions (3) and (6).

\item
To calculate the strain energy density of
a network of self-avoiding chains, and to
find adjustable parameters in the governing
equations by fitting experimental data.
\end{enumerate}
To derive constitutive equations, we employ
the Boltzmann formula, according to which
the probability $p_{0}({\bf Q})$
that a chain has an end-to-end vector ${\bf Q}$
is expressed in terms of the configurational free energy
$U_{0}({\bf Q})$ as
\begin{equation}
p_{0}({\bf Q})=\exp \Bigl (
-\frac{U_{0}({\bf Q})}{k_{\rm B}T}\Bigr ).
\end{equation}
The function $U_{0}({\bf Q})$ determines the energy
of a chain in the reference (stress-free) state.
When macro-deformation is applied to a polymer
network,
the initial end-to-end vector ${\bf Q}$ of the chain
is transformed into some vector ${\bf q}$
at an instant $t\geq 0$ (time $t=0$ corresponds
to the application of external loads).
For an affine network of chains,
the distribution function $p(t,{\bf q})$
of the vector ${\bf q}$ obeys an appropriate
Smoluchowski equation \cite{DE86} that is solved
explicitly.

Given $p(t,{\bf q})$, the configurational free energy
of a chain in the actual (deformed) state $U(t,{\bf q})$
is described by the equation similar to Eq. (11),
\begin{equation}
p(t,{\bf q})=\exp \Bigl (
-\frac{U(t,{\bf q})}{k_{\rm B}T}\Bigr ).
\end{equation}
When the functions $U_{0}({\bf Q})$ and $U(t,{\bf q})$
are found, the increment of the configurational free
energy is given by
\begin{equation}
\Delta U=U-U_{0}.
\end{equation}
The strain energy per chain $W(t)$ is determined
by averaging the increment $\Delta U$ with respect to
an appropriate distribution function ($p_{0}$ or
$p$).
A general expression for $W$ will be developed in Section 3.
At volume-preserving deformation
of a Gaussian chain with the distribution function
\begin{equation}
p_{0}({\bf Q})=\Bigl (\frac{3}{2\pi b^{2}}\Bigr )^{\frac{3}{2}}
\exp \Bigl (-\frac{3Q^{2}}{2b^{2}}\Bigr ),
\end{equation}
where $Q=|{\bf Q}|$, this formula is transformed into Eq. (1),
when the averaging is performed with respect to
the current distribution of end-to-end vectors,
and it implies the Mooney--Rivlin constitutive law
\begin{equation}
W = \frac{1}{2} k_{\rm B}T \Bigl [ a (J_{1}-3)
+(1-a) ( J_{2}-3) \Bigr ],
\end{equation}
when the averaging is performed with respect
to the functions $p$ and $p_{0}$ with the weights
$a$ and $1-a$, respectively.

Regrettably, the strain energy of a chain cannot
be expressed as an elementary function of principal
stretches for an arbitrary three-dimensional
deformation.
To develop such an analytical expression, we apply
an approximation procedure similar to the three-chain
hypothesis \cite{JG43}, according to which
``the network of chains\ldots
is mathematically equivalent to three independent
sets of chains respectively parallel to the axes
of a three-dimensional Cartesian coordinate system"
\cite{TR79}.

By using the three-chain approximation, explicit
formulas are derived for the strain energy of
self-avoiding chains with distribution functions (3) and (6).
An important conclusion of our analysis is
that for an incompressible network,
the strain energy of a chain with the
stretched exponential distribution function
(3) is described (up to a small correction term)
by the three-term Odgen law \cite{Ogd84}
\begin{equation}
W=\frac{1}{2}k_{\rm B}T\sum_{n=1}^{3}\kappa_{n}
\sum_{m=1}^{3} \Bigl [ a (\lambda_{m}^{\frac{\alpha_{n}}{2}}
-3)+(1-a)(\lambda_{m}^{-\frac{\alpha_{n}}{2}} -3)\Bigr ].
\end{equation}
Unlike the original Ogden formula \cite{Ogd84},
where $a$, $\alpha_{n}$ and $\kappa_{n}$ ($n=1,2,3$)
are treated as adjustable parameters,
our expression contains only 2 material
constants, while the exponents $\alpha_{n}$ equal
$1$, $3$ and $5$, respectively.
This result provides a micro-mechanical basis for
the Ogden model, on the one hand, and allows the
number of experimental constants in Eq. (16)
to be reduced noticeably, on the other.

For the des Cloizeaux law (6), an expression for
the strain energy per chain $W$ is developed
for an arbitrary (positive and negative)
exponent $\alpha$.
The function $W$ differs from that described by Eq. (16)
by an additional term that depends on $\alpha$.
Surprisingly, approximation of experimental data
reveals that the best fit of observations is
reached at negative values of $\alpha$.
As these values have no physical meaning within
the model of self-avoiding random walks (although
they are not forbidden from the physical standpoint),
we conclude that the Ogden constitutive law (16)
describes the elastic response of all chains
that can be modeled as self-avoiding random walks.

Finally, the mechanical response of self-avoiding
chains is analyzed at small deformations.
At uniaxial tension (compression) with small
strains, the elastic behavior of a chain is
entirely described by the only parameter, an
analog of the Young's modulus.
The elastic modulus is proportional
to $k_{\rm B}T$, but the coefficient of
proportionality depends on the distribution
function $p_{0}({\bf Q})$.
It is found that self-avoiding chains with the
stretched exponential distribution function (3)
are ``stronger" than Gaussian chains with the
distribution function (14) by a factor of three,
whereas the ratio of the Young's modulus of
a self-avoiding chain with the des Cloizeaux
distribution function to that of a Gaussian
chain linearly increases with exponent $\alpha$.

The exposition is organized as follows.
The Smoluchowski equation for the distribution
function $p(t,{\bf q})$ is resolved in Section 2
for an arbitrary time-dependent deformation of
an affine network.
A formula for the strain-energy of a chain is developed
is Section 3.
The strain energy of a Gaussian chain is calculated
in Section 4.
A three-chain approximation procedure is introduced
in Section 5.
The mechanical energy of a self-avoiding chain with
the stretched-exponential distribution function
is found in Section 6.
This expression is applied in Section 7
to fit experimental data on several elastomers.
The strain energy of a chain whose statistics is
governed by the des Cloizeaux law (6) is determined
in Section 8, and the corresponding constitutive law
is validated in Section 9.
The elastic moduli of self-avoiding chains are calculated
in Section 10.
Some concluding remarks are formulated in Section 11.
To avoid technical details in the main text,
necessary calculations are given in Appendices.

\section{Transformation of the distribution function}

With begin with the solution of the Smoluchowski
equation for the distribution function of end-to-end
vectors $p$.
Consider a chain with an end-to-end vector ${\bf Q}$
in the reference state and an end-to-end vector
${\bf q}$ is the actual state at time $t\geq 0$.
In an affine network,
transformation of the reference state of the chain
into its deformed state is described by the formula
\begin{equation}
{\bf q}={\bf F}(t)\cdot {\bf Q},
\end{equation}
where ${\bf F}$ stands for the deformation
gradient for macro-deformation (at this stage
of the analysis, we do not impose the incompressibility
condition on the tensor ${\bf F}$).
The function ${\bf F}(t)$ obeys the differential
equation
\begin{equation}
\frac{d{\bf F}}{dt} ={\bf L}\cdot {\bf F},
\qquad
{\bf F}(0)={\bf I},
\end{equation}
where ${\bf L}(t)$ is the velocity gradient,
${\bf I}$ is the unit tensor.

The distribution function $p(t,{\bf q})$ of the end-to-end
vector ${\bf q}$ satisfies the equation \cite{DE86}
\begin{equation}
\frac{\partial p}{\partial t}
= -\frac{\partial}{\partial {\bf q}}
\cdot \Bigl ( {\bf L}\cdot {\bf q} p\Bigr )
= -{\cal I}_{1}({\bf D})p
-\frac{\partial p}{\partial {\bf q}}
\cdot {\bf L}\cdot {\bf q} ,
\end{equation}
with the initial condition
\begin{equation}
p(0,{\bf q})=p_{0}({\bf q}).
\end{equation}
Here
\[
{\bf D}=\frac{1}{2}\Bigl ( {\bf L}
+{\bf L}^{\top}\Bigr )
\]
is the rate-of-strain tensor, $\top$ denotes
transpose, and ${\cal I}_{m}$ stands for the
$m$th principal invariant of a tensor ($m=1,2,3$).
Simple algebra implies that the solution of Eqs. (19)
and (20) reads
\begin{equation}
p(t,{\bf q})= p_{0}({\bf F}^{-1}(t)\cdot {\bf q})
\exp \Bigl [-\int_{0}^{t} {\cal I}_{1}\Bigl (
{\bf D}(s)\Bigr )ds\Bigr ].
\end{equation}
It follows from Eq. (18) that the third
principal invariant of the deformation gradient
obeys the equation
\begin{equation}
\frac{d{\cal I}_{3}({\bf F})}{dt}
={\cal I}_{1}({\bf D})
{\cal I}_{3}({\bf F}).
\end{equation}
The solution of Eq. (22) with the initial
condition ${\cal I}_{3}({\bf F}(0))=1$
reads
\begin{equation}
{\cal I}_{3}({\bf F}(t))=
\exp \Bigl [ \int_{0}^{t}
{\cal I}_{1}\Bigl ({\bf D}(s)\Bigr )
ds \Bigr ].
\end{equation}
Equations (21) and (23) imply that
\begin{equation}
p(t,{\bf q})= \frac{p_{0}({\bf F}^{-1}(t)
\cdot {\bf q})}{{\cal I}_{3}({\bf F}(t))}.
\end{equation}
To demonstrate that function (24) satisfies the
normalization condition
\begin{equation}
\int p(t,{\bf q})d{\bf q}=1,
\end{equation}
we substitute expression (24) into the left-hand
side of Eq. (25) and find that
\[
\int p(t,{\bf q})d{\bf q}=
\frac{1}{{\cal I}_{3}({\bf F}(t))}
\int p_{0}({\bf F}^{-1}(t) \cdot
{\bf q}) d{\bf q}.
\]
Introducing the new variable ${\bf Q}$
by Eq. (17) and bearing in mind that
\begin{equation}
d{\bf q}={\cal I}_{3}({\bf F}) d{\bf Q},
\end{equation}
we arrive at the formula
\[
\int p(t,{\bf q})d{\bf q}=
\int p_{0}({\bf Q}) d{\bf Q}.
\]
Equation (25) follows from this equality and Eq. (4).

\section{Strain energy of a chain}

Our aim now is to calculate the strain energy of
a chain in an affine network by using Eq. (24).
The configurational free energies of a chain
in the reference and actual states are connected with
appropriate distribution functions by Eqs. (11) and (12),
\begin{equation}
U_{0}({\bf Q})=-k_{\rm B}T \ln p_{0}({\bf Q}),
\qquad
U(t,{\bf q})=-k_{\rm B}T \ln p(t,{\bf q}).
\end{equation}
There are two ways to determine the strain energy per
strand $W$.
According to the first,
the increment $\Delta U$ of the configurational
free energy caused by transition from the
reference state to the actual state,
\[
\Delta U(t,{\bf Q})=U(t,{\bf Q})-U_{0}({\bf Q})
=-k_{\rm B}T \Bigl [ \ln p(t,{\bf Q})
-\ln p_{0}({\bf Q})\Bigr ]
\]
is averaged with the help of the distribution function
in the reference state,
\begin{equation}
W_{1}(t) = -k_{\rm B}T \int \Bigl [
\ln p(t,{\bf Q})-\ln p_{0}({\bf Q})\Bigr ]
p_{0}({\bf Q}) d{\bf Q}.
\end{equation}
According to the other approach,
we calculate the increment of the configurational
free energy with respect to the actual state,
\[
\Delta U(t,{\bf q})=U_{0}({\bf q})-U(t,{\bf q})
=-k_{\rm B}T \Bigl [ \ln p_{0}({\bf q})-
\ln p(t,{\bf q})\Bigr ],
\]
and average it by using the distribution
function in the deformed state,
\begin{equation}
W_{2}(t) = -k_{\rm B}T \int \Bigl [
\ln p_{0}({\bf q})-\ln p(t, {\bf q})\Bigr ]
p(t,{\bf q}) d{\bf q}.
\end{equation}
We begin with transformation of Eq. (28).
Substituting expression (24) into this equality
and using Eq. (26), we obtain
\begin{equation}
W_{1}(t)
= -k_{\rm B}T \biggl \{ \int \Bigl [
\ln p_{0}({\bf F}^{-1}(t)\cdot {\bf Q})
-\ln p_{0}({\bf Q})\Bigr ]p_{0}({\bf Q})d{\bf Q}
-\ln {\cal I}_{3}({\bf F}(t)) \biggr \}.
\end{equation}
We now proceed with Eq. (29), combine this equality with
Eq. (24), we find that
\[
W_{2}(t) = \frac{k_{\rm B}T}{{\cal I}_{3}({\bf F}(t))}
\int \Bigl [ \ln p_{0}({\bf F}^{-1}(t)\cdot {\bf q})
-\ln p_{0}({\bf q})
-\ln {\cal I}_{3} ({\bf F}(t))\Bigr ]
p_{0}({\bf F}^{-1}(t) \cdot {\bf q}) d{\bf q}.
\]
Introducing the new variable ${\bf Q}$ by
Eq. (17) and using Eqs. (25) and (26), we find that
\begin{equation}
W_{2}(t)= k_{\rm B}T
\biggl \{ \int \Bigl [ \ln p_{0} ({\bf Q})
-\ln p_{0}({\bf F}(t)\cdot {\bf Q})
\Bigr ] p_{0}({\bf Q}) d{\bf Q}
-\ln {\cal I}_{3}({\bf F}(t)) \biggr \}.
\end{equation}
It seems natural to define the strain
energy per strand $W$ as the weighted sum of
the strain energies $W_{1}$ and $W_{2}$
that are determined by employing different ways
of averaging of the configurational free energy,
\begin{equation}
W=(1-a) W_{1}+a W_{2},
\end{equation}
where $a\in [0,1]$ is a material parameter.
Combining Eqs. (30)--(32), we arrive at the formula
\begin{eqnarray}
W(t) &=& k_{\rm B}T \biggl \{-(2a-1) \ln {\cal I}_{3}({\bf F}(t))
+\int \biggl [ a \Bigl ( \ln p_{0} ({\bf Q})
-\ln p_{0}({\bf F}(t)\cdot {\bf Q})
\Bigr )
\nonumber\\
&&+(1-a)\Bigl (\ln p_{0}({\bf Q})
-\ln p_{0}({\bf F}^{-1}(t)\cdot {\bf Q})
\Bigr )\biggr ]p_{0}({\bf Q})d{\bf Q} \biggr \}.
\end{eqnarray}
To proceed with transformations of Eq. (33),
an additional hypothesis is needed regarding
the function $p_{0}({\bf Q})$.
Assuming the end-to-end vectors to be distributed
uniformly in the reference state, we set
\begin{equation}
p_{0}({\bf Q})=P(Q^{2}),
\end{equation}
where $Q^{2}={\bf Q}\cdot {\bf Q}$,
and $P(r)$ is a given function of a scalar argument $r$.
Substituting expression (34) into Eq. (33)
and taking into account that
\begin{eqnarray*}
&& \Bigl ({\bf F}\cdot {\bf Q})\cdot
\Bigl ({\bf F}\cdot {\bf Q}\Bigr )
={\bf Q}\cdot {\bf F}^{\top}
\cdot {\bf F}\cdot {\bf Q}
={\bf Q}\cdot {\bf C}\cdot {\bf Q},
\nonumber\\
={\bf Q}\cdot {\bf B}^{-1}\cdot {\bf Q},
\nonumber\\
&& {\cal I}_{3}^{2} ({\bf F})=J_{3},
\end{eqnarray*}
where the left and right Cauchy--Green
deformation tensors are determined by
the conventional formulas
\begin{equation}
{\bf B}={\bf F}\cdot {\bf F}^{\top},
\qquad
{\bf C}={\bf F}^{\top} \cdot {\bf F},
\end{equation}
we find that
\begin{eqnarray}
W &=& k_{\rm B}T \biggl \{
-\frac{1}{2} (2a-1) \ln J_{3}
+\int \biggl [ a \Bigl (\ln P (Q^{2})
-\ln P({\bf Q}\cdot {\bf C}\cdot {\bf Q})\Bigr )
\nonumber\\
&&+(1-a)\Bigl ( \ln P(Q^{2}-\ln P({\bf Q}\cdot
{\bf B}^{-1}\cdot {\bf Q})
\Bigr )\biggr ] P(Q^{2})d{\bf Q} \biggr \}.
\end{eqnarray}
It is easy to show (Appendix A) that
\begin{equation}
\int \ln P({\bf Q}\cdot {\bf B}^{-1}
\cdot {\bf Q}) P(Q^{2})d{\bf Q}
=\int \ln P({\bf Q}\cdot {\bf C}^{-1}
\cdot {\bf Q}) P(Q^{2})d{\bf Q}.
\end{equation}
It follows from Eqs. (36) and (37) that the strain energy
per strand reads
\begin{eqnarray}
W &=& k_{\rm B}T \biggl \{
-\frac{1}{2} (2a-1) \ln J_{3}
+ \int \biggl [ a \Bigl (\ln P (Q^{2})
-\ln P({\bf Q}\cdot {\bf C}\cdot {\bf Q})\Bigr )
\nonumber\\
&&+(1-a)\Bigl ( \ln P(Q^{2})-\ln P({\bf Q}\cdot
{\bf C}^{-1}\cdot {\bf Q})
\Bigr )\biggr ] P(Q^{2})d{\bf Q}\biggr \}.
\end{eqnarray}
Our aim now is to apply Eq. (38) in order to determine
the strain energy of a Gaussian chain with
distribution function (14).

\section{Strain energy of a Gaussian chain}

According to Eqs. (14) and (34), the function $P(r)$
reads
\[
P(r)=\Bigl (\frac{3}{2\pi b^{2}}\Bigr )^{\frac{3}{2}}
\exp \Bigl (-\frac{3r}{2b^{2}}\Bigr ).
\]
Simple algebra (Appendix B) implies that
\begin{eqnarray}
&& \int \Bigl ( \ln P(Q^{2})
-\ln P({\bf Q}\cdot {\bf C}
\cdot{\bf Q})\Bigr )P(Q^{2})d{\bf Q}
=\frac{1}{2} (J_{1}-3),
\nonumber\\
&& \int \Bigl (\ln P(Q^{2})
-\ln P({\bf Q}\cdot {\bf C}^{-1}
\cdot{\bf Q})\Bigr )
P(Q^{2})d{\bf Q}
=\frac{1}{2} ( J_{-1}-3 ),
\end{eqnarray}
where
$J_{-m}={\cal I}_{m}({\bf C}^{-1})$ ($m=1,2,3$).
Substitution of expressions (39) into Eq. (38)
implies that
\begin{equation}
W = \frac{1}{2} k_{\rm B}T \Bigl [ a ( J_{1}-3)
+(1-a) ( J_{-1}-3)
-(2a-1) \ln J_{3} \Bigr ].
\end{equation}
Bearing in mind that $J_{-1}=J_{2}/J_{3}$,
we present Eq. (40) in the form
\begin{equation}
W = \frac{1}{2} k_{\rm B}T \Bigl [ a ( J_{1}-3)
+(1-a) ( \frac{J_{2}}{J_{3}}-3)
-(2a-1) \ln J_{3} \Bigr ].
\end{equation}
Equation (41) implies that for an incompressible
polymer network with $J_{3}=1$,
the strain energy of a chain is given by
the Moony--Rivlin law (15).
Equation (41) is reduced to the constitutive
relation (1) for an incompressible neo--Hookean
medium (which is traditionally associated with
the response of Gaussian chains) at $a=1$ only
(when the averaging of the increment of
configurational free energy $\Delta U$ is
performed with respect to the distribution
function $p$ in the actual state).
It is worth noting that our results are in accord
with those developed in \cite{KS00,HK97} by using
another approach (a tube model for rubbery polymers).

\section{Three-chain approximation}

Our aim now is to introduce a three-chain approximation
for the strain energy per chain based on expression (40)
for the mechanical energy of a Gaussian chain.
Bearing in mind that
\[
J_{1}=\lambda_{1}
+\lambda_{2}
+\lambda_{3},
\qquad
J_{-1}=\frac{1}{\lambda_{1}}
+\frac{1}{\lambda_{2}}
+\frac{1}{\lambda_{3}},
\qquad
J_{3}=\lambda_{1}
\lambda_{2}
\lambda_{3},
\]
where $\lambda_{m}$ ($m=1,2,3$) are eigenvalues
of the Cauchy--Green tensor ${\bf C}$,
we find from Eq. (40) that
\begin{equation}
W = \frac{1}{2} k_{\rm B}T \Bigl [
a ( \lambda_{1}
+\lambda_{2}
+\lambda_{3}-3 )
+(1-a) \Bigl ( \frac{1}{\lambda_{1}}
+\frac{1}{\lambda_{2}}
+\frac{1}{\lambda_{3}}-3\Bigr )
-(2a-1) \ln (\lambda_{1}
\lambda_{2}
\lambda_{3}) \Bigr ].
\end{equation}
Let us consider now three networks of chains.
The principal axes of the Cauchy--Green
deformation tensor ${\bf C}^{(m)}$
for the $m$th network ($m=1,2,3$)
coincide with the eigenvectors of the tensor ${\bf C}$,
whereas the corresponding eigenvalues $\{ \lambda_{1}^{(m)},
\lambda_{2}^{(m)}, \lambda_{3}^{(m)} \}$
are given by
\begin{equation}
\begin{array}{lll}
\lambda_{1}^{(1)}=\lambda_{1},
&
\lambda_{2}^{(1)}=1,
&
\lambda_{3}^{(1)}=1,
\\
\lambda_{1}^{(2)}=1,
&
\lambda_{2}^{(2)}=\lambda_{2},
&
\lambda_{3}^{(2)}=1,
\\
\lambda_{1}^{(3)}=1,
&
\lambda_{2}^{(3)}=1,
&
\lambda_{3}^{(3)}=\lambda_{3}.
\end{array}
\end{equation}
It follows from Eqs. (42) and (43) that the strain energy
per chain in the $m$th network reads
\begin{equation}
W^{(m)} = \frac{1}{2} k_{\rm B}T \Bigl [
a ( \lambda_{m}-1 )
+(1-a) \Bigl ( \frac{1}{\lambda_{m}}
-1\Bigr )
-(2a-1) \ln \lambda_{m} \Bigr ].
\end{equation}
Equations (42) and (44) imply that
\begin{equation}
W=\sum_{m=1}^{3} W^{(m)} .
\end{equation}
Formula (45) may be referred to as the
three-chain approximation of the strain
energy $W$ (it slightly differs from the three-chain
approach proposed in \cite{JG43}).
Equation (45) provides an exact expression for
the strain energy of a Gaussian chain.
For non-Gaussian chains, this formula
may serve as a convenient approximation for
the strain energy.

Returning to general expression (38) and
applying the three-chain approximation, we
find that the strain energy $W$ is given
by Eq. (45), where $W^{(m)}$ read
\begin{eqnarray}
W^{(m)} &=& k_{\rm B}T \biggl \{
-\frac{1}{2} (2a-1) \ln \lambda_{m}
+ \int \biggl [
a \Bigl (\ln P (Q^{2})
-\ln P({\bf Q}\cdot
{\bf C}^{(m)}\cdot {\bf Q})\Bigr )
\nonumber\\
&&+(1-a)\Bigl ( \ln P(Q^{2})
-\ln P({\bf Q}\cdot
({\bf C}^{(m)})^{-1}\cdot {\bf Q})
\Bigr )\biggr ] P(Q^{2})d{\bf Q}\biggr \}.
\end{eqnarray}
We now choose a spherical coordinate frame $\{ Q,
\theta,\phi \}$, whose axes coincide with the
eigenvectors of the tensor ${\bf C}^{(m)}$
and the $z$-axis is directed along the eigenvector
with the eigenvalue that differs from unity.
The quadratic forms
${\bf Q}\cdot{\bf C}^{(m)}\cdot {\bf Q}$
and
${\bf Q}\cdot({\bf C}^{(m)})^{-1}\cdot {\bf Q}$
are calculated as follows:
\[
{\bf Q}\cdot{\bf C}^{(m)}\cdot {\bf Q}
=Q^{2}(\sin^{2}\theta+\lambda_{m}\cos^{2}\theta),
\quad
{\bf Q}\cdot({\bf C}^{(m)})^{-1}\cdot {\bf Q}
=Q^{2}(\sin^{2}\theta+\frac{1}{\lambda_{m}}\cos^{2}\theta).
\]
Substituting these expressions into Eq. (46)
and performing integration over $\phi$, we find that
\begin{eqnarray*}
W^{(m)} &=& k_{\rm B}T \biggl \{
-\frac{1}{2} (2a-1) \ln \lambda_{m}
+ 2\pi \int_{0}^{\infty}  P(Q^{2})Q^{2} d Q
\int_{0}^{\pi} \sin \theta d\theta
\nonumber\\
&& \times \biggl [
a \Bigl (\ln P (Q^{2})
-\ln P\Bigl ( Q^{2}
(\sin^{2}\theta+\lambda_{m}\cos^{2}\theta)\Bigr )
\nonumber\\
&&+(1-a)\Bigl ( \ln P(Q^{2})
-\ln P\Bigl ( Q^{2}
(\sin^{2}\theta+\lambda_{m}^{-1}\cos^{2}\theta)\Bigr )
\biggr ] \biggr \}.
\end{eqnarray*}
Introducing the new variable $z=\cos\theta$, we obtain
\begin{eqnarray}
W^{(m)} &=& k_{\rm B}T \biggl \{
-\frac{1}{2} (2a-1) \ln \lambda_{m}
+ 4\pi \int_{0}^{\infty}  P(Q^{2})Q^{2} d Q
\nonumber\\
&& \times
\int_{0}^{1} \biggl [
a \Bigl (\ln P (Q^{2})
-\ln P\Bigl ( Q^{2} (1
+(\lambda_{m}-1) z^{2})\Bigr )
\nonumber\\
&&+(1-a)\Bigl ( \ln P(Q^{2})
-\ln P\Bigl ( Q^{2}
(1+(\lambda_{m}^{-1}-1)z^{2})\Bigr )
\biggr ] dz \biggr \}.
\end{eqnarray}
Our aim now is to calculate the strain energy $W$
of a self-avoiding chain with distribution
function (3) by using Eqs. (45) and (47).

\section{The stretched exponential distribution function}

Equations Eqs. (3) and (34) imply that
\begin{equation}
P(r)=p^{0}\exp \Bigl [
-\beta \Bigl (\frac{r}{R^{2}}\Bigr )^{\delta}\Bigr ],
\end{equation}
where the pre-factor $p^{0}$ is given by Eq. (5).
Substitution of Eq. (48) into Eqs. (45) and (47) results
in (see Appendix C for detail)
\begin{eqnarray}
W &=& k_{\rm B}T \biggl \{
-\frac{1}{2} (2a-1) \ln (
\lambda_{1}\lambda_{2}\lambda_{3})
+\biggl [ a\sum_{m=1}^{3}
\Bigl (\frac{1}{6}\lambda_{m}^{\frac{5}{2}}
+\frac{5}{24}\lambda_{m}^{\frac{3}{2}}
+\frac{5}{16}\lambda_{m}^{\frac{1}{2}}\Bigr )
\nonumber\\
&&+(1-a)\sum_{m=1}^{3} \Bigl (\frac{1}{6}
\lambda_{m}^{-\frac{5}{2}}
+\frac{5}{24}\lambda_{m}^{-\frac{3}{2}}
+\frac{5}{16}\lambda_{m}^{-\frac{1}{2}}\Bigr )
-\frac{33}{16} \biggr ]+\tilde{W}\biggr \},
\end{eqnarray}
where
\begin{equation}
\tilde{W}=\frac{5}{16} \biggl [
\sum_{m=1}^{3} \Bigl (a G(\lambda_{m})
+(1-a)G(\lambda_{m}^{-1})\Bigr )
-3 \biggr ].
\end{equation}
The function $G(z)$ in Eq. (50) reads
\begin{equation}
G(z)=\left \{\begin{array}{ll}
(1-z)^{-\frac{1}{2}}\arcsin (1-z)^{\frac{1}{2}}, & z<1,
\\
1, & z=1,\\
(z-1)^{-\frac{1}{2}} \ln (z^{\frac{1}{2}}
+(z-1)^{\frac{1}{2}}),  & z>1.
\end{array}
\right .
\end{equation}
This function does not exceed $\frac{\pi}{2}$
for any $z\geq 0$, and it monotonically decreases
(however, rather weakly) with $z$ and vanishes
as $z\to \infty$.

At finite strains, when the norm of the Cauchy--Green
tensor ${\bf C}$ is large compared with unity,
the last term in Eq. (49) may be disregarded.
Under this assumption, Eq. (49) is reduced to the
constitutive equation of the Ogden medium with
a special choice of material parameters.
\begin{eqnarray*}
W_{0} &=& k_{\rm B}T \biggl [
-\frac{1}{2} (2a-1) \ln J_{3}
+ a \sum_{m=1}^{3} \Bigl (
\frac{1}{6}\lambda_{m}^{\frac{5}{2}}
+\frac{5}{24}\lambda_{m}^{\frac{3}{2}}
+\frac{5}{16}\lambda_{m}^{\frac{1}{2}}
-\frac{11}{16} \Bigr )
\nonumber\\
&&+(1-a)\sum_{m=1}^{3} \Bigl (\frac{1}{6}
\lambda_{m}^{-\frac{5}{2}}
+\frac{5}{24}\lambda_{m}^{-\frac{3}{2}}
+\frac{5}{16}\lambda_{m}^{-\frac{1}{2}}
-\frac{11}{16} \Bigr ) \biggr ].
\end{eqnarray*}
For an incompressible network, the first term in
this equality vanishes, and we obtain, in accord
with Eq. (16),
\begin{eqnarray}
W_{0} &=&  k_{\rm B}T \biggl [
a \sum_{m=1}^{3} \Bigl (
\frac{1}{6}\lambda_{m}^{\frac{5}{2}}
+\frac{5}{24}\lambda_{m}^{\frac{3}{2}}
+\frac{5}{16}\lambda_{m}^{\frac{1}{2}}
-\frac{11}{16} \Bigr )
\nonumber\\
&&+(1-a)\sum_{m=1}^{3} \Bigl (\frac{1}{6}
\lambda_{m}^{-\frac{5}{2}}
+\frac{5}{24}\lambda_{m}^{-\frac{3}{2}}
+\frac{5}{16}\lambda_{m}^{-\frac{1}{2}}
-\frac{11}{16} \Bigr ) \biggr ].
\end{eqnarray}
To demonstrate that Eq. (52) provides quite an
acceptable approximation of the strain energy $W$
for conventional loading programs,
we calculate the strain energies
determined by Eqs. (49) and (52) for uniaxial
extension of an incompressible network,
\begin{equation}
\lambda_{1}=k^{2},
\qquad
\lambda_{2}=k^{-1},
\qquad
\lambda_{3}=k^{-1},
\end{equation}
where $k$ stands for elongation ratio,
and for simple shear with
\[
\lambda_{1}=1,
\quad
\lambda_{2}=\Bigl (1+\frac{k^{2}}{2}\Bigr )
+\sqrt{\Bigl (1+\frac{k^{2}}{2}\Bigr )^{2}-1},
\quad
\lambda_{3}=\Bigl (1+\frac{k^{2}}{2}\Bigr )
-\sqrt{\Bigl (1+\frac{k^{2}}{2}\Bigr )^{2}-1},
\]
where $k$ denotes shear.
For definiteness, we set $a=0.5$.
The quantities $W$ and $W_{0}$
are plotted versus $k$ in Figures 1 and 2.
These figures show that the dimensionless strain
energies $\overline{W}=W/(k_{\rm B}T)$ practically
coincide when they are calculated with and
without the correction term $\tilde{W}$.

Formula (52) determines the strain energy of
an individual chain.
Neglecting the energy of interaction between chains
(this energy is conventionally accounted for
by the incompressibility condition \cite{DE86}),
we calculate the strain energy per unit volume
of a network as the sum of strain energies of chains,
\begin{equation}
w_{0} =  \mu_{1}
\sum_{m=1}^{3} \Bigl (\frac{1}{6}\lambda_{m}^{\frac{5}{2}}
+\frac{5}{24}\lambda_{m}^{\frac{3}{2}}
+\frac{5}{16}\lambda_{m}^{\frac{1}{2}}
-\frac{11}{16} \Bigr )
+\mu_{2} \sum_{m=1}^{3} \Bigl (\frac{1}{6}
\lambda_{m}^{-\frac{5}{2}}
+\frac{5}{24}\lambda_{m}^{-\frac{3}{2}}
+\frac{5}{16}\lambda_{m}^{-\frac{1}{2}}
-\frac{11}{16} \Biggr ),
\end{equation}
where
\begin{equation}
\mu_{1}=k_{\rm B}a MT,
\qquad
\mu_{2}=k_{\rm B}(1-a) MT,
\end{equation}
and $M$ is the number of chains per unit volume.
Equation (54) involves only two coefficients,
$\mu_{1}$ and $\mu_{2}$, to be found by fitting
observations.
Our aim now is to show that constitutive law (54)
correctly describes experimental data at uniaxial
tension, uniaxial compression and equi-biaxial
tension of elastomers.

\section{Results of numerical simulation}

At three-dimensional deformation of an
incompressible medium without rotations,
the principal Cauchy stresses $\sigma_{m}$
are expressed in terms of the principal
stretches $\lambda_{m}$ by the formulas
\begin{equation}
\sigma_{m}=\lambda_{m}\frac{\partial w}{\partial
\lambda_{m}}-\tilde{p},
\end{equation}
where $\tilde{p}$ is an unknown pressure,
and $w$ is a strain energy per unit volume.
Substitution of expression (54) into Eq. (56) results in
\begin{equation}
\sigma_{m}=\lambda_{m} \Bigl [ \mu_{1}F_{0}(\lambda_{m})
-\frac{\mu_{2}}{\lambda_{m}^{2}}F_{0}(\frac{1}{\lambda_{m}})
\Bigr ]-\tilde{p},
\end{equation}
where
\begin{equation}
F_{0}(\lambda)=\frac{5}{4} \Bigl (
\frac{1}{3}\lambda^{\frac{3}{2}}
+\frac{1}{4}\lambda^{\frac{1}{2}}
+\frac{1}{8}\lambda^{-\frac{1}{2}}\Bigr ).
\end{equation}
For uniaxial extension with an elongation ratio
$k$, the principal stretches are given by Eq. (53).
Excluding the pressure $\tilde{p}$ from the condition
$\sigma_{2}=\sigma_{3}=0$, we find from
Eq. (57) that
the engineering stress $\sigma_{\rm e}=\sigma_{1}/k$
is given by
\begin{equation}
\sigma_{\rm e}=\frac{1}{k} \Bigl [
\mu_{1}\Bigl ( k^{2} F_{0}(k^{2})
-\frac{1}{k}F_{0}(\frac{1}{k})\Bigr )
+\mu_{2} \Bigl ( kF_{0}(k)
-\frac{1}{k^{2}}F_{0}(\frac{1}{k^{2}})\Bigr )
\Bigr ].
\end{equation}
For equi-biaxial deformation of an incompressible
material, the principal stretches $\lambda_{m}$
read
\begin{equation}
\lambda_{1}=k^{2},
\qquad
\lambda_{2}=k^{2},
\qquad
\lambda_{3}=k^{-4},
\end{equation}
where $k$ stands for elongation ratio.
We substitute Eq. (60) into Eq. (57),
exclude the unknown pressure $\tilde{p}$ from the
condition $\sigma_{3}=0$, and find the
Cauchy stresses $\sigma_{1}=\sigma_{2}=\sigma$.
The engineering stress $\sigma_{\rm e}=\sigma/k$
is determined as
\begin{equation}
\sigma_{\rm e}=\frac{1}{k} \Bigl [ \mu_{1}
\Bigl ( k^{2}F_{0}(k^{2})
-\frac{1}{k^{4}}F_{0}(\frac{1}{k^{4}})\Bigr )
+\mu_{2} \Bigl ( k^{4}F_{0}(k^{4})
-\frac{1}{k^{2}}F_{0}(\frac{1}{k^{2}})
\Bigr ) \Bigr ].
\end{equation}
Given an experimental dependence $\sigma_{\rm e}(k)$,
the coefficients $\mu_{1}$ and $\mu_{2}$ in Eqs. (59)
and (61) are found by the least-squares technique
from the condition of minimum of the function
\[
R=\sum_{k_{n}} \Bigl [ \sigma_{\rm e}^{\rm exp}(k_{n})
-\sigma_{\rm e}^{\rm num}(k_{n})\Bigr ]^{2},
\]
where the sum is calculated over all elongation ratios
$k_{n}$ at which measurements are reported,
$\sigma_{\rm e}^{\rm exp}$ is the stress
measured in a test, and $\sigma_{\rm e}^{\rm num}$
is given by Eqs. (59) and (61).

We begin with fitting the observations on uniaxial
extension of natural rubber with various amounts of
cross-linker tetramethylthiuram disulfide (TMTD)
reported by Kl\"{u}ppel and Schramm \cite{KS00}.
The experimental data together with their approximation
by Eqs. (58) and (59) are depicted in Figure 3
(where phr means parts per hundred parts of rubber).
This figure shows that Eq. (59) with the only parameter
$\mu_{1}$ (we set $\mu_{2}=0$ to reduce the number
of material constants)
provides good agreement between the observations
and the results of numerical simulation.
The elastic modulus $\mu_{1}$ is plotted versus
concentration of cross-linker $\phi$ in Figure 4,
where the experimental data are approximated by
the linear dependence
\begin{equation}
\mu_{1}=\mu_{1}^{(1)}\phi .
\end{equation}
The coefficient $\mu_{1}^{(1)}$ in Eq. (62) is
determined by the least-squares technique.
Formula (62) confirms our assumption that the
energy of interaction between chains in a network
may be taken into account with the help of the
incompressibility condition.
It implies that the modulus $\mu_{1}$ is proportional
to the amount of cross-linker, which, in turn, is
proportional to the number of chains per unit volume $M$,
in accord with Eq. (55).

We proceed with matching the experimental data on
uniaxial compression of carbon black-filled
chloroprene rubber reported by Bergstr\"{o}m and
Boyce \cite{BB98}.
The experimental dependencies $\sigma_{\rm e}(k)$
measured at various amounts of filler are depicted in
Figure 5 together with their approximations by Eq. (59)
with $\mu_{2}=0$.
The modulus $\mu_{1}$ is plotted versus concentration
of filler $\phi$ in Figure 6.
The experimental data for $\mu_{1}(\phi)$ are
approximated by the linear dependence
\begin{equation}
\mu_{1}=\mu_{1}^{(0)}+\mu_{1}^{(1)}\phi,
\end{equation}
where the coefficients $\mu_{1}^{(0)}$ and
$\mu_{1}^{(1)}$ are found by the least-squares
method.
Figure 6 demonstrates that Eq. (63), which is
conventionally employed in the mechanics of
composites to describe the effect of filler
on the shear modulus, correctly predicts the
experimental data.

Figures 3 and 5 show that the observations
on natural and chloroprene rubbers at uniaxial
tension and compression can be approximated
by the model with $a=1$, i.e. when averaging
of the configurational free energy is performed
by using the distribution function of end-to-end vectors
in the deformed state.
To demonstrate that this is not the rule in
the general case, we fit experimental data
on uniaxial tension--compression of carbon
black-reinforced natural rubber reported
by Hartmann et al. \cite{HTS03}
and of polybutadiene rubber presented by Roland et al.
\cite{RMH99}.
The observations and the results of numerical analysis
are depicted in Figures 7 and 8, which reveal good
agreement between the experimental data and the results
of simulation.

To validate the governing equations, we approximate
the observations on uniaxial extension of a synthetic
rubber reported by Chevalier et al. \cite{CCH01},
find adjustable parameters $\mu_{1}$ and $\mu_{2}$,
calculate the engineering stress $\sigma_{\rm e}$ at
equi-biaxial tension by Eq. (61),
and compare the results of numerical simulation
with the experimental data.
Figure 9 demonstrates that the model provides quite an
acceptable prediction of the elastic response at
equi-biaxial deformation with moderate finite
strains, $1.0 \leq k<1.4$.

Figures 3 to 9 reveal that Eq. (54) with
two adjustable parameters can correctly describe
the mechanical response of elastomers at large
deformations.
It is worth noting that this assertion is not trivial
(the fact that the Ogden model adequately describes
observations has been confirmed by numerous experimental
data in the past three decades), because the number
of material constants in Eq. (54) is noticeably smaller
than that in the original expression (16) for the
strain energy density.

Another result to be mentioned is that
the description of some experimental data
by Eq. (54) is not perfect.
As examples, we refer to the observations depicted
in Figures 7 and 9.
This conclusion may be attributed to the fact that Eq. (3)
for the stretched-exponential distribution function of
a self-avoiding random walk is overly simplified,
and more sophisticated formula (6) should be used.
Our aim now is to develop governing equations for a
network of self-avoiding chains with the des Cloizeaux
distribution function of end-to-end vectors.

\section{The des Cloizeaux distribution function}

For a SARW with the distribution function (6),
the function $P(r)$ reads
\begin{equation}
P(r)=p^{0}\Bigl (\frac{r}{R^{2}}\Bigr )^{\alpha}
\exp \Bigl [-\beta \Bigl (\frac{r}{R^{2}}\Bigr )^{\delta}
\Bigr ],
\end{equation}
where $\delta$ is given by Eq. (10),
and $p^{0}$ is determined by Eq. (7).
Substitution of expression (64) into Eqs. (45) and (47)
implies that (see Appendix D for detail)
\begin{eqnarray}
W &=&  k_{\rm B}T \biggl \{
-(2a-1) (\alpha+\frac{1}{2})
\ln (\lambda_{1}\lambda_{2}\lambda_{3})
\nonumber\\
&&+\Bigl (1+\frac{2}{3}\alpha\Bigr ) \biggl [
a \sum_{m=1}^{3}
\Bigl (\frac{1}{6}\lambda_{m}^{\frac{5}{2}}
+\frac{5}{24}\lambda_{m}^{\frac{3}{2}}
+\frac{5}{16}\lambda_{m}^{\frac{1}{2}}-\frac{11}{16}\Bigr )
\nonumber\\
&&+(1-a)\sum_{m=1}^{3} \Bigl (\frac{1}{6}
\lambda_{m}^{-\frac{5}{2}}
+\frac{5}{24}\lambda_{m}^{-\frac{3}{2}}
+\frac{5}{16}\lambda_{m}^{-\frac{1}{2}}
-\frac{11}{16} \Bigr )\biggr ]+\tilde{W}
\nonumber\\
&& -2 \alpha \sum_{m=1}^{3} \Bigl (aH(\lambda_{m})
+(1-a)H(\lambda_{m}^{-1})\Bigr )\biggr \},
\end{eqnarray}
where
\begin{equation}
\begin{array}{ll}
H(z)=\frac{\arctan \sqrt{z-1}}{\sqrt{z-1}}-1,
& z>1,\\
H(z)=0, & z=1,\\
H(z)=\frac{1}{2\sqrt{1-z}}\ln\frac{1+\sqrt{1-z}}{1-\sqrt{1-z}}-1,
& z<1.
\end{array}
\end{equation}
The function $H(z)$ monotonically decreases in $(0,\infty)$
and tends to $-1$ when $z\to \infty$.
The graph of this function is presented in Figure 10.

Using notation (52) and neglecting the small
term $\tilde{W}$ compared with $W_{0}$,
we find from Eq. (65) that
\begin{equation}
W = \Bigl (1+\frac{2}{3}\alpha\Bigr ) W_{0}
- k_{\rm B}T \biggl [ (2a-1) (\alpha+\frac{1}{2})\ln J_{3}
+2\alpha \sum_{m=1}^{3} \Bigl (aH(\lambda_{m})
+(1-a)H(\lambda_{m}^{-1})\Bigr )\biggr ].
\end{equation}
In the tensor form, Eq. (67) reads (see Appendix E)
\begin{eqnarray}
\frac{W}{k_{\rm B}T} &=&
-(2a-1) (\alpha+\frac{1}{2})\ln J_{3}
+\Bigl (1+\frac{2}{3}\alpha\Bigr )
\biggl [
a \Bigl (\frac{1}{6}{\cal I}_{1}({\bf C}_{\rm e}^{\frac{5}{2}})
+\frac{5}{24} {\cal I}_{1}({\bf C}_{\rm e}^{\frac{3}{2}})
+\frac{5}{16}{\cal I}_{1}({\bf C}_{\rm e}^{\frac{1}{2}})\Bigr )
\nonumber\\
&&+(1-a)
\Bigl (\frac{1}{6}{\cal I}_{1}({\bf C}_{\rm e}^{-\frac{5}{2}})
+\frac{5}{24} {\cal I}_{1}({\bf C}_{\rm e}^{-\frac{3}{2}})
+\frac{5}{16}{\cal I}_{1}({\bf C}_{\rm e}^{-\frac{1}{2}})\Bigr )
-\frac{33}{16}\biggr ]
- \alpha \sum_{k=1}^{\infty} \frac{2^{k+1}}{2k+1}
\nonumber\\
&&\times
\Bigl [a (-1)^{k} {\cal I}_{1}({\bf E}_{\rm C}^{k})
+(1-a){\cal I}_{1}({\bf E}_{\rm A}^{k})\Bigr ],
\end{eqnarray}
where
\begin{equation}
{\bf E}_{\rm C}=\frac{1}{2} ({\bf C}-{\bf I}),
\qquad
{\bf E}_{\rm A}=\frac{1}{2} ({\bf I}-{\bf C}^{-1})
\end{equation}
are the Cauchy and Almansi strain tensors,
respectively.

At volume-preserving deformations, Eqs. (52) and (67)
imply that
\begin{eqnarray*}
W &=& k_{\rm B}T \biggl \{
a \sum_{m=1}^{3} \Bigl [ (1+\frac{2}{3}\alpha)
\Bigl (
\frac{1}{6}\lambda_{m}^{\frac{5}{2}}
+\frac{5}{24}\lambda_{m}^{\frac{3}{2}}
+\frac{5}{16}\lambda_{m}^{\frac{1}{2}}
-\frac{11}{16} \Bigr )
-2\alpha H(\lambda_{m})\Bigr ]
\nonumber\\
&&+(1-a)\sum_{m=1}^{3} \Bigl [ (1+\frac{2}{3}\alpha)
\Bigl (\frac{1}{6}
\lambda_{m}^{-\frac{5}{2}}
+\frac{5}{24}\lambda_{m}^{-\frac{3}{2}}
+\frac{5}{16}\lambda_{m}^{-\frac{1}{2}}
-\frac{11}{16} \Bigr ) -
2\alpha H(\lambda_{m}^{-1}) \Bigr ]
\biggr \}.
\end{eqnarray*}
Applying the same approach that has been employed
in Section 6, that is neglecting the energy
of interaction between chains
and assuming the strain energy of a network
to coincide with the sum of strain energies
of individual chains, we arrive at the formula
for the strain energy density per unit volume of
an incompressible network
\begin{eqnarray}
w &=& a \sum_{m=1}^{3} \Bigl [ \mu_{1} \Bigl (
\frac{1}{6}\lambda_{m}^{\frac{5}{2}}
+\frac{5}{24}\lambda_{m}^{\frac{3}{2}}
+\frac{5}{16}\lambda_{m}^{\frac{1}{2}}
-\frac{11}{16} \Bigr )
-\mu_{2} H(\lambda_{m})\Bigr ]
\nonumber\\
&&+(1-a)\sum_{m=1}^{3} \Bigl [ \mu_{1}
\Bigl (\frac{1}{6}
\lambda_{m}^{-\frac{5}{2}}
+\frac{5}{24}\lambda_{m}^{-\frac{3}{2}}
+\frac{5}{16}\lambda_{m}^{-\frac{1}{2}}
-\frac{11}{16} \Bigr )
-\mu_{2} H(\lambda_{m}^{-1}) \Bigr ],
\end{eqnarray}
where we set
\begin{equation}
\mu_{1}=k_{\rm B}T (1+\frac{2}{3}\alpha)M,
\qquad
\mu_{2}=2k_{\rm B}T \alpha M .
\end{equation}
In the general case, constitutive law (70) involves
3 adjustable parameters, $a$, $\mu_{1}$ and $\mu_{2}$.
It is worth noting a substantial difference between
the parameter $\mu_{2}$ in Eqs. (55) and (70).
According to the physical meaning of the quantity
$\mu_{2}$ in Eq. (55), this parameter should always
be positive.
On the contrary, the parameter $\mu_{2}$ is Eq. (70)
may be positive [this corresponds to the case $\alpha>0$
in Eq. (6)], as well as negative [when
$\alpha\in (-\frac{3}{2},0)$].
The latter condition ensures that the integral of the
function $P(Q^{2})$ over the entire space converges.

Our aim now is to develop stress--strain relations
for finite deformations of an incompressible
network,
to derive explicit expressions for the principal
Cauchy stresses,
and to compare experimental data on elastomers
with results of numerical analysis.

\section{Fitting of observations}

It follows from Eq. (70) that
\begin{equation}
\frac{\partial w}{\partial \lambda_{m}}=
a\Bigl [ \mu_{1}F_{0}(\lambda_{m})+\mu_{2}F(\lambda_{m})
\Bigr ]
-\frac{1-a}{\lambda_{m}^{2}}
\Bigl [\mu_{1}F_{0}(\frac{1}{\lambda_{m}})
+\mu_{2}F(\frac{1}{\lambda_{m}})\Bigr ],
\end{equation}
where $F_{0}(z)$ is given by Eq. (58), and
\begin{equation}
F(z)=-\frac{dH}{dz}(z)
=\left \{\begin{array}{ll}
\frac{1}{2(z-1)} (
\frac{\arctan\sqrt{z-1}}{\sqrt{z-1}}
-\frac{1}{z}), & z>1,
\\
\frac{1}{3}, & z=1,
\\
\frac{1}{2(1-z)}( \frac{1}{z}
-\frac{1}{2\sqrt{1-z}}
\ln\frac{1+\sqrt{1-z}}{1-\sqrt{1-z}}),
& z<1.
\end{array}\right .
\end{equation}
The function $F(z)$ is positive and continuous
at any $z>0$,
it monotonically decreases with $z$ and vanishes
when $z\to\infty$.
The graph of the function $F(z)$ is depicted in Figure 10.

Substitution of expression (72) into Eq. (57) results
in the formula
\begin{equation}
\sigma_{m} = -p+ \mu_{1} \lambda_{m} \Bigl [a F_{0}(\lambda_{m})
-\frac{1-a}{\lambda_{m}^{2}}F_{0}(\frac{1}{\lambda_{m}})
\Bigr ]
+\mu_{2} \lambda_{m} \Bigl [a F(\lambda_{m})
-\frac{1-a}{\lambda_{m}^{2}}F(\frac{1}{\lambda_{m}})
\Bigr ].
\end{equation}
According to Eq. (74), at uniaxial extension
with an elongation ratio $k$, see Eq. (53),
the engineering stress $\sigma_{\rm e}$ is given by
\begin{eqnarray}
\sigma_{\rm e} &=& \frac{1}{k}\biggl \{
\mu_{1} \Bigl [
a\Bigl ( k^{2}F_{0}(k^{2})
-\frac{1}{k}F_{0}(\frac{1}{k})\Bigr )
+(1-a)\Bigl ( kF_{0}(k)
-\frac{1}{k^{2}}F_{0}(\frac{1}{k^{2}})\Bigr )\Bigr ]
\nonumber\\
&&+\mu_{2}\Bigl [
a\Bigl ( k^{2}F(k^{2})-\frac{1}{k}F(\frac{1}{k})\Bigr )
+(1-a)\Bigl ( k F(k) -\frac{1}{k^{2}}F(\frac{1}{k^{2}})
\Bigr )\Bigr ]\biggr \}.
\end{eqnarray}
At equi-biaxial extension with an elongation ratio $k$,
see Eq. (60), the engineering stress reads
\begin{eqnarray}
\sigma_{\rm e} &=& \frac{1}{k}\biggl \{\mu_{1}
\Bigl [ a\Bigl ( k^{2}F_{0}(k^{2})
-\frac{1}{k^{4}}F_{0}(\frac{1}{k^{4}})\Bigr )
+(1-a)\Bigl ( k^{4}F_{0}(k^{4})
-\frac{1}{k^{2}}F_{0}(\frac{1}{k^{2}})\Bigr ) \Bigr ]
\nonumber\\
&&+\mu_{2}\Bigl [ a\Bigl ( k^{2}F(k^{2})
-\frac{1}{k^{4}}F(\frac{1}{k^{4}})\Bigr )
+(1-a)\Bigl ( k^{4}F(k^{4})
-\frac{1}{k^{2}}F(\frac{1}{k^{2}})\Bigr ) \Bigr ]\biggr \}.
\end{eqnarray}
At biaxial extension with elongation ratios $k_{1}$ and $k_{2}$,
\begin{equation}
\lambda_{1}=k_{1}^{2},
\qquad
\lambda_{2}=k_{2}^{2},
\qquad
\lambda_{3}=(k_{1}k_{2})^{-2},
\end{equation}
the engineering stresses $\sigma_{{\rm e},1}=\sigma_{1}/k_{1}$
and $\sigma_{{\rm e},2}=\sigma_{2}/k_{2}$ are determined by
\begin{eqnarray}
\sigma_{{\rm e},1} &=& \frac{1}{k_{1}}\biggl \{\mu_{1}
\Bigl [ a\Bigl ( k_{1}^{2}F_{0}(k_{1}^{2})
-\frac{1}{k_{1}^{2}k_{2}^{2}}F_{0}(\frac{1}{k_{1}^{2}k_{2}^{2}})\Bigr )
+(1-a)\Bigl ( k_{1}^{2}k_{2}^{2}F_{0}(k_{1}^{2}k_{2}^{2})
-\frac{1}{k_{1}^{2}}F_{0}(\frac{1}{k_{1}^{2}})\Bigr ) \Bigr ]
\nonumber\\
&&+\mu_{2}\Bigl [ a\Bigl ( k_{1}^{2}F(k_{1}^{2})
-\frac{1}{k_{1}^{2}k_{2}^{2}}F(\frac{1}{k_{1}^{2}k_{2}^{2}})\Bigr )
+(1-a)\Bigl ( k_{1}^{2}k_{2}^{2}F(k_{1}^{2}k_{2}^{2})
-\frac{1}{k_{1}^{2}}F(\frac{1}{k_{1}^{2}})\Bigr ) \Bigr ]
\biggr \},
\nonumber\\
\sigma_{{\rm e},2} &=& \frac{1}{k_{2}}\biggl \{\mu_{1}
\Bigl [ a\Bigl ( k_{2}^{2}F_{0}(k_{2}^{2})
-\frac{1}{k_{1}^{2}k_{2}^{2}}F_{0}(\frac{1}{k_{1}^{2}k_{2}^{2}})\Bigr )
+(1-a)\Bigl ( k_{1}^{2}k_{2}^{2}F_{0}(k_{1}^{2}k_{2}^{2})
-\frac{1}{k_{2}^{2}}F_{0}(\frac{1}{k_{2}^{2}})\Bigr ) \Bigr ]
\nonumber\\
&&+\mu_{2}\Bigl [ a\Bigl ( k_{2}^{2}F(k_{2}^{2})
-\frac{1}{k_{1}^{2}k_{2}^{2}}F(\frac{1}{k_{1}^{2}k_{2}^{2}})\Bigr )
+(1-a)\Bigl ( k_{1}^{2}k_{2}^{2}F(k_{1}^{2}k_{2}^{2})
-\frac{1}{k_{2}^{2}}F(\frac{1}{k_{2}^{2}})\Bigr ) \Bigr ]
\biggr \}.
\end{eqnarray}
Our purpose now is to apply Eqs. (75), (76) and (78)
in order to approximate observations at uniaxial,
equi-biaxial and biaxial extension of elastomers.
To find the constants $a$, $\mu_{1}$ and $\mu_{2}$,
we divide the interval $[0,1]$, where the parameter
$a$ is located, into $I$ subintervals by
the points $a^{(i)}=i\Delta a$
($i=1,\ldots,I-1$) with $\Delta a=1/I$.
Given $a^{(i)}$,
the coefficients $\mu_{1}$ and $\mu_{2}$ are found
by the least-squares method from the condition of minimum
of the function $R$.
The ``best-fit" parameter $a$ is determined from
the condition of minimum of this function
on the set $ \{ a^{(i)} \}$.

We begin with the experimental data on uniaxial
tension--compression of two types of carbon
black-filled natural rubber.
The experimental stress--strain curves reported
by Roland et al. \cite{RMH99} are depicted
in Figure 11 together with their approximations
by Eq. (75).
This figure shows that Eq. (75) with $a=1$
(averaging with respect to the distribution function
in the actual state) and $\mu_{2}<0$ (a weakly-singular
at $Q=0$ des Cloizeaux distribution function)
ensures good agreement with the observations.

The latter peculiarity of the model appear to
be rather typical.
To confirm this assertion, we approximate
the experimental data on equi-biaxial extension
of natural rubber reported by James at el. \cite{JGS75}.
The observations together with their fit are
plotted in Figure 12.
It is worth noting some similarity between the
material constants determined by matching the
experimental data for three different types of
natural rubber: the modulus $\mu_{1}$ is of
order of $(2\div 4)\cdot 10^{-2}$ MPa, and the
modulus $|\mu_{2}|$ is of order of $3\div 7$ MPa.

To validate the model, we analyze observations on
biaxial extension of poly(dimethyl\-sil\-oxane) network
(volume fraction of PDMS $\phi=0.463$) reported by
Kawamura et al. \cite{KUK02}.
The adjustable parameters are found by matching the
dependence $\sigma_{{\rm e},2}(k_{2})$ at $k_{1}=1.9$.
Afterwards, the functions $\sigma_{{\rm e},2}(k_{2})$
at $k_{1}=1.1$, 1.3, 1.5 and 1.7 are predicted
numerically, and the results of simulation are compared
with the experimental data.
Figure 13 shows excellent agreement between the
observations and the model predictions.

\section{The elastic modulus of self-avoiding chains}

We return now to expressions (52) and (70)
for the strain energy of an incompressible
network of self-avoiding chains in order to
assess the effect of excluded volume on the
elastic modulus per chain at small deformations.

For a network of Gaussian chains under uniaxial
tension with an elongation ratio $k$, the engineering
stress $\sigma_{\rm e}$ is given by the conventional
formula \cite{Tre75}
\[
\sigma_{\rm e}=(\mu_{1}+\mu_{2}k^{-1})(k-k^{-2}),
\]
where
\begin{equation}
\mu_{1}=\frac{1}{2} k_{\rm B}aTM,
\qquad
\mu_{2}=\frac{1}{2} k_{\rm B}(1-a)TM.
\end{equation}
At small strains, when $k=1+\epsilon$ with $\epsilon\ll 1$,
we obtain
$\sigma_{\rm e}=3(\mu_{1}+\mu_{2})\epsilon $.
The Young's modulus of a network $E$ is given by
\begin{equation}
E=\frac{\sigma_{\rm e}}{\epsilon}.
\end{equation}
Substituting expressions (79) into Eq. (80)
and introducing the Young's modulus per chain
as
\begin{equation}
E_{\rm c}=\frac{E}{M},
\end{equation}
we arrive at the formula
\begin{equation}
E_{\rm c}=\frac{3}{2}k_{\rm B} T.
\end{equation}
For a network with strain energy density (52),
the engineering stress at uniaxial extension
is given by Eqs. (58) and (59).
Omitting simple algebra, we find
that at small strains,
\[
\sigma_{\rm e}=\frac{305}{64}(\mu_{1}+\mu_{2})\epsilon .
\]
It follows from this equality and Eqs. (55), (80),
and (81) that
\[
E_{\rm c}=\frac{305}{64}k_{\rm B}T.
\]
Combining Eq. (82) and this equation, we arrive at the formula
\begin{equation}
\frac{E_{\rm c}^{\rm Ogden}}{E_{\rm c}^{\rm Gauss}}
=\frac{305}{96}\approx 3.1771,
\end{equation}
which means that the ``rigidity" of a self-avoiding
chain exceeds that of a Gaussian chain by
a factor of three (approximately).
A reason for the anesthetic ratio in Eq. (83)
is that the contribution of $\tilde{W}$
into the strain energy density $W$ is neglected.

We proceed with the evaluation of the Young's
modulus for a chain whose distribution function is
described by des Cloizeaux law (6).
It follows from Eq. (73) that
\[
\frac{dF}{dz}(1)=-\frac{2}{5}.
\]
Using this expression, we find from Eqs. (58) and (75)
that at uniaxial extension with small strains,
\[
E=\frac{305}{64}\mu_{1}
-\frac{1}{5}\mu_{2}.
\]
Substitution of expressions (71) into this formula
results in
\[
E_{\rm c}=\Bigl ( \frac{305}{64}+\frac{1333}{480}\alpha
\Bigr )k_{\rm B}T.
\]
It follows from this equality and Eqs. (82) and (83) that
\begin{equation}
\frac{E_{\rm c}^{\rm des Cloizeaux}}{E_{\rm c}^{\rm Gauss}}
=\frac{E_{\rm c}^{\rm Ogden}}{E_{\rm c}^{\rm Gauss}}
+\frac{1333}{720} \alpha
\approx 3.1771 +1.8514\alpha .
\end{equation}
Equations (83) and (84) imply that the Young's
modulus per chain $E_{\rm c}$ is independent
of the characteristic length $R$ and
the exponent $\beta$ in Eqs. (3) and (6),
and it increases with $\alpha$ in Eq. (6)
being proportional to $2\alpha$ (approximately).

Qualitatively, the latter conclusion is not surprising.
The growth of $\alpha$ is tantamount
to the fact that the tail of distribution
function (6) becomes thicker, while the latter
implies (within the conventional ``force--stretch"
approach in the statistical mechanics of polymers)
an enhancement of the rigidity of chains.

To assess the Young's modulus found in our
approximation of the observations, we use Eq. (71),
which implies that
\[
\alpha=-\frac{1}{2}\Bigl (\frac{1}{3}-\frac{\mu_{1}}{\mu_{2}}
\Bigr )^{-1}.
\]
Using the results presented in Figures 11 to 13,
we find that $\alpha=-1.46$ (Figure 11) and $-1.48$
(Figure 12) for natural rubber, and $-1.49$
(Figure 13) for PDMS.
For these parameters, Eq. (84) implies that the ratio
$E_{\rm c}^{\rm des Cloizeaux}/E_{\rm c}^{\rm Gauss}$
is close to 0.43 for PDMS and is located between
0.47 and 0.49 for natural rubber.
The difference between these quantities and the ratio
of elastic moduli predicted by Eq. (83) may provide
an explanation, why constitutive law (52)
does not provide an adequate approximation of
the observations depicted in Figures 11 to 13.

\section{Concluding remarks}

A general expression has been derived for
the strain energy of a polymer chain with
an arbitrary distribution function of
end-to-end vectors.
This formula has been applied to determine
the strain energy of a chain that is modeled
as a self-avoiding random walk.
By using three-chain approximation method,
we has found that the strain energy of a network
of chains with the stretched exponential distribution
function coincides with the Ogden law with
a special choice of adjustable parameters.
In this case, the stress--strain relations
involve only two material constants that
are found by matching experimental data at
uniaxial tension, uniaxial compression
and equi-biaxial tension of elastomers.
Good agreement is demonstrated between
the observations and the results of numerical
simulation.

For the generalized stretched-exponential
distribution function (the des Cloizeaux law),
a new expression is developed for the strain
energy density of a network.
In this case, the governing equations contain
three material constants to be found by matching
observations.
It is shown that the stress--strain relations
provide quite an acceptable fit of observations
at uniaxial and bi-axial extension of filled
elastomers even when the Ogden relations poorly
approximate the experimental data.
It is demonstrated that the latter case corresponds
to rubber-like materials, whose elastic modulus
at uniaxial extension is lower than that for Gaussian
chains.
\newpage
\appendix

\section*{Appendix A}
\renewcommand{\theequation}{A-\arabic{equation}}
\setcounter{equation}{0}

To derive Eq. (37), we, first, choose a spherical
coordinate frame $\{ Q,\theta,\phi \}$,
whose axes coincide with the eigenvectors of
the symmetric tensor ${\bf B}^{-1}$.
In the spherical coordinates, the quadratic form
${\bf Q}\cdot {\bf B}^{-1}\cdot {\bf Q}$
is given by
\[
{\bf Q}\cdot {\bf B}^{-1}\cdot {\bf Q}
=Q^{2} \Bigl [ (B^{-1})_{1}\sin^{2}\theta\cos^{2}\phi
+(B^{-1})_{2}\sin^{2}\theta\sin^{2}\phi
+(B^{-1})_{3}\cos^{2}\theta \Bigr ],
\]
where $(B^{-1})_{m}$ ($m=1,2,3$) are eigenvalues
of the tensor ${\bf B}^{-1}$.
The integral on the left-hand side of Eq. (37) reads
\begin{eqnarray*}
\int \ln P({\bf Q}\cdot {\bf B}^{-1}
\cdot {\bf Q}) P(Q^{2})d{\bf Q}
&=& \int_{0}^{\infty} P(Q^{2}) Q^{2} dQ
\int_{0}^{\pi} \sin \theta d\theta
\int_{0}^{2\pi} \Bigl [ (B^{-1})_{1}
\sin^{2}\theta\cos^{2}\phi
\nonumber\\
&& +(B^{-1})_{2}\sin^{2}\theta\sin^{2}\phi
+(B^{-1})_{3}\cos^{2}\theta \Bigr ] d\phi.
\end{eqnarray*}
We now choose the spherical coordinate frame
$\{ Q^{\prime},\theta^{\prime},\phi^{\prime} \}$,
whose axes coincide with the eigenvectors of
the symmetric tensor ${\bf C}^{-1}$.
Repeating the above calculations, we find
the integral on the right-hand side of Eq. (37),
\begin{eqnarray*}
\int \ln P({\bf Q}\cdot {\bf C}^{-1}
\cdot {\bf Q}) P(Q^{2})d{\bf Q}
&=& \int_{0}^{\infty} P(Q^{\prime 2}) Q^{\prime 2} dQ^{\prime}
\int_{0}^{\pi} \sin \theta^{\prime} d\theta^{\prime}
\int_{0}^{2\pi} \Bigl [ (C^{-1})_{1}
\sin^{2}\theta^{\prime} \cos^{2}\phi^{\prime}
\nonumber\\
&& +(C^{-1})_{2}\sin^{2}\theta^{\prime}\sin^{2}\phi^{\prime}
+(C^{-1})_{3}\cos^{2}\theta^{\prime} \Bigr ] d\phi^{\prime},
\end{eqnarray*}
where $(C^{-1})_{m}$ ($m=1,2,3$) are eigenvalues
of the tensor ${\bf C}^{-1}$.
Bearing in mind that the eigenvalues of the Cauchy-Green
tensors ${\bf B}^{-1}$ and ${\bf C}^{-1}$
coincide, we conclude from these relations that
Eq. (37) is satisfied.

\section*{Appendix B}
\renewcommand{\theequation}{B-\arabic{equation}}
\setcounter{equation}{0}

We begin with transformation of the term
\begin{equation}
\int \ln P({\bf Q}\cdot {\bf C}
\cdot{\bf Q}) P(Q^{2})d{\bf Q}
=\frac{3}{2}\ln \frac{3}{2\pi b^{2}}
-\frac{3}{2b^{2}}
\Bigl (\frac{3}{2\pi b^{2}}\Bigr )^{\frac{3}{2}}
\int ({\bf Q}\cdot {\bf C}\cdot {\bf Q})
\exp \Bigl (-\frac{3Q^{2}}{2b^{2}}\Bigr ) d{\bf Q}.
\end{equation}
In spherical coordinates $\{ Q,\theta,\phi \}$
directed along the eigenvectors of the symmetrical
tensor ${\bf C}$,
the integral in Eq. (B-1) is presented as
\begin{eqnarray*}
\int ({\bf Q}\cdot {\bf C}\cdot {\bf Q})
\exp \Bigl (-\frac{3Q^{2}}{2b^{2}}\Bigr ) d{\bf Q}
&=& \int_{0}^{\infty} Q^{4} \exp
\Bigl (-\frac{3Q^{2}}{2b^{2}}\Bigr ) d Q
\int_{0}^{\pi} \sin \theta d\theta
\int_{0}^{2\pi} \Bigl [
\lambda_{1}\sin^{2}\theta\cos^{2}\phi
\nonumber\\
&& +\lambda_{2}\sin^{2}\theta\sin^{2}\phi
+\lambda_{3}\cos^{2}\theta \Bigr ] d\phi,
\end{eqnarray*}
where $\lambda_{m}$ ($m=1,2,3$) are eigenvalues
of the tensor ${\bf C}$.
The integrals on the right-hand side of this equality
are calculated explicitly,
\begin{eqnarray*}
&& \int_{0}^{\infty} Q^{4} \exp
\Bigl (-\frac{3Q^{2}}{2b^{2}}\Bigr ) d Q
=3\Bigl (\frac{\pi}{2} \Bigr )^{\frac{1}{2}}
\Bigl (\frac{b^{2}}{3} \Bigr )^{\frac{5}{2}},
\\
&& \int_{0}^{\pi} \sin \theta d\theta
\int_{0}^{2\pi} \Bigl [ \lambda_{1}
\sin^{2}\theta\cos^{2}\phi
+\lambda_{2}\sin^{2}\theta\sin^{2}\phi
+\lambda_{3}\cos^{2}\theta \Bigr ] d\phi
=\frac{4\pi}{3} J_{1}.
\end{eqnarray*}
Substitution of these expressions into Eq. (B-1)
results in
\begin{equation}
\int \ln P({\bf Q}\cdot {\bf C}
\cdot{\bf Q}) P(Q^{2})d{\bf Q}
=\frac{3}{2}\ln \frac{3}{2\pi b^{2}}
-\frac{1}{2} J_{1}.
\end{equation}
Bearing in mind that
$P(Q^{2})=P({\bf Q}\cdot
{\bf I}\cdot {\bf Q})$,
we find from this equality that
\begin{equation}
\int \ln P(Q^{2}) P(Q^{2})d{\bf Q}
=\frac{3}{2}\ln \frac{3}{2\pi b^{2}}
-\frac{3}{2}.
\end{equation}
Equation (39) follows from Eqs. (B-2) and (B-3).

\section*{Appendix C}
\renewcommand{\theequation}{C-\arabic{equation}}
\setcounter{equation}{0}

Substitution of Eq. (48) into Eq. (47) implies that
\begin{eqnarray}
W^{(m)} &=& k_{\rm B}T \biggl \{
-\frac{1}{2} (2a-1) \ln \lambda_{m}
+ K \biggl [ \int_{0}^{1} \biggl (a
(1+(\lambda_{m}-1)z^{2})^{\delta}
\nonumber\\
&& +(1-a)(1+(\lambda_{m}^{-1}-1)z^{2})^{\delta}
\biggr )dz-1\biggr ]\biggr \},
\end{eqnarray}
where
\[
K=4\pi \beta P_{0}
\int_{0}^{\infty} \exp \Bigl [-\beta \Bigl (
\frac{Q}{R}\Bigr )^{2\delta}\Bigr ]
\Bigl (\frac{Q}{R}\Bigr )^{2\delta} Q^{2} d Q.
\]
Combining this equality with Eq. (5) and introducing
the new variable $r=Q/R$, we find that
\[
K=\beta \frac{\int_{0}^{\infty} \exp(-\beta r^{3}) r^{5}dr}
{\int_{0}^{\infty} \exp(-\beta r^{3}) r^{2}dr}.
\]
We set $z=\beta^{\frac{1}{3}}r$ in this equality
and obtain
\[
K=\frac{\int_{0}^{\infty} \exp(-z^{3}) z^{5}dz}
{\int_{0}^{\infty} \exp(-z^{3}) z^{2}dz}.
\]
The ratio of the integrals is easily calculated by
integration by parts,
\begin{equation}
K=1.
\end{equation}
Substitution of expression (C-2) into Eq. (C-1)
results in
\begin{eqnarray}
W^{(m)} &=& k_{\rm B}T \biggl \{
-\frac{1}{2} (2a-1) \ln \lambda_{m}
+ \biggl [ \int_{0}^{1} \biggl (a
(1+(\lambda_{m}-1)z^{2})^{\delta}
\nonumber\\
&& +(1-a)(1+(\lambda_{m}^{-1}-1)z^{2})^{\delta}
\biggr )dz-1\biggr ]\biggr \}.
\end{eqnarray}
To find an analytical expression for the
function $W^{(m)}$, it is necessary to calculate
the integral
\begin{equation}
L(\mu,\delta)=\int_{0}^{1} (1+\mu z^{2})^{\delta} dz
\end{equation}
for $\mu=\lambda_{m}-1$ and
$\mu=\lambda_{m}^{-1}-1$, and
for $\delta$ given by Eq. (10).
We begin with the case $\mu >0$.
Integration by parts implies that
\[
\int (1+\mu z^{2})^{\frac{5}{2}} dz=\frac{1}{6}
\Bigl [ z(1+\mu z^{2})^{\frac{5}{2}}
+5\int (1+\mu z^{2})^{\frac{3}{2}}dz \Bigr ].
\]
It follows from this formula that
\begin{equation}
L(\mu,\frac{5}{2})=\frac{1}{6} \Bigl [ (1+\mu)^{\frac{5}{2}}
+5L(\mu,\frac{3}{2})\Bigr ],
\end{equation}
where the function $L$ is given by Eq. (C-4).
Applying integration by parts to the new integral,
we obtain
\[
\int (1+\mu z^{2})^{\frac{3}{2}} dz=\frac{1}{4}
\Bigl [ z(1+\mu z^{2})^{\frac{3}{2}}
+3\int (1+\mu z^{2})^{\frac{1}{2}}dz \Bigr ],
\]
which results in
\begin{equation}
L(\mu,\frac{3}{2})=\frac{1}{4} \Bigl [ (1+\mu)^{\frac{3}{2}}
+3L(\mu,\frac{1}{2})\Bigr ].
\end{equation}
It is easy to check that
\[
\int (1+\mu z^{2})^{\frac{1}{2}} dz
=\frac{1}{2} \Bigl [ z\sqrt{1+\mu z^{2}}
+\frac{1}{\sqrt{\mu}}\ln (\sqrt{\mu}z+\sqrt{1+\mu z^{2}})
\Bigr ].
\]
It follows from this equality that
\begin{equation}
L(\mu,\frac{1}{2})=\frac{1}{2}\Bigl [
\sqrt{1+\mu}+\frac{1}{\sqrt{\mu}}\ln (\sqrt{\mu}
+\sqrt{1+\mu})\Bigr ].
\end{equation}
Combination of Eqs. (C-5) to (C-7) yields
\begin{equation}
L(\mu,\frac{5}{2})=\frac{1}{6}(1+\mu)^{\frac{5}{2}}
+\frac{5}{24} (1+\mu)^{\frac{3}{2}}
+\frac{5}{16}(1+\mu)^{\frac{1}{2}}
+\frac{5}{16\sqrt{\mu}}\ln (\sqrt{\mu}
+\sqrt{1+\mu}).
\end{equation}
Returning to the initial notation, we find from
Eqs. (C-4) and (C-8) that
\begin{eqnarray}
\int_{0}^{1} (1+(\lambda_{m}-1)z^{2})^{\frac{5}{2}}dz
&=& \frac{1}{6}\lambda_{m}^{\frac{5}{2}}
+\frac{5}{24}\lambda_{m}^{\frac{3}{2}}
+\frac{5}{16}\lambda_{m}^{\frac{1}{2}}
+\frac{5}{16} (\lambda_{m}-1)^{-\frac{1}{2}}
\nonumber\\
&&\times
\ln \Bigl (\lambda_{m}^{\frac{1}{2}}
+(\lambda_{m}-1)^{\frac{1}{2}}\Bigr )
\qquad
(\lambda_{m}>1).
\end{eqnarray}
We now consider the case $\mu<0$, set $\mu_{1}=-\mu$,
and calculate the integral
\[
L_{1}(\mu_{1},\frac{5}{2})=
\int_{0}^{1} (1-\mu_{1} z^{2})^{\frac{5}{2}} dz
\qquad
(\mu_{1}>0).
\]
Repeating the above transformations, we find that
\begin{eqnarray}
L_{1}(\mu_{1},\frac{5}{2}) &=& \frac{1}{6}\Bigl [
(1-\mu_{1})^{\frac{5}{2}}+5L_{1}(\mu_{1},\frac{3}{2})
\Bigr ],
\nonumber\\
L_{1}(\mu_{1},\frac{3}{2}) &=& \frac{1}{4}\Bigl [
(1-\mu_{1})^{\frac{3}{2}}+3L_{1}(\mu_{1},\frac{1}{2})
\Bigr ],
\nonumber\\
L_{1}(\mu_{1},\frac{1}{2}) &=& \frac{1}{2}\Bigl [
(1-\mu_{1})^{\frac{1}{2}}+\frac{1}{\sqrt{\mu_{1}}}
\arcsin \sqrt{\mu_{1}}\Bigr ].
\end{eqnarray}
To develop the latter equality, we use the formula
\[
\int (1-\mu_{1} z^{2})^{\frac{1}{2}}dz=\frac{1}{2}
\Bigl [ z \sqrt{1-\mu_{1}z^{2}}
+\frac{1}{\mu_{1}}\arcsin \sqrt{\mu_{1}}z \Bigr ] .
\]
It follows from Eq. (C-10) that
\begin{equation}
L_{1}(\mu_{1},\frac{5}{2})=\frac{1}{6}(1-\mu_{1})^{\frac{5}{2}}
+\frac{5}{24} (1-\mu_{1})^{\frac{3}{2}}
+\frac{5}{16}(1-\mu_{1})^{\frac{1}{2}}
+\frac{5}{16\sqrt{\mu_{1}}}\arcsin \sqrt{\mu_{1}}.
\end{equation}
Bearing in mind that $\mu_{1}=1-\lambda_{m}$,
we find from Eq. (C-11) that
\begin{eqnarray}
\int_{0}^{1} (1+(\lambda_{m}-1)z^{2})^{\frac{5}{2}}dz
&=& \frac{1}{6}\lambda_{m}^{\frac{5}{2}}
+\frac{5}{24}\lambda_{m}^{\frac{3}{2}}
+\frac{5}{16}\lambda_{m}^{\frac{1}{2}}
+\frac{5}{16} (1-\lambda_{m})^{-\frac{1}{2}}
\nonumber\\
&&\times
\arcsin (1-\lambda_{m})^{\frac{1}{2}}
\qquad
(\lambda_{m}<1).
\end{eqnarray}
It is obvious that
\begin{eqnarray}
\int_{0}^{1} (1+(\lambda_{m}-1)z^{2})^{\frac{5}{2}}dz
&=& \frac{1}{6}\lambda_{m}^{\frac{5}{2}}
+\frac{5}{24}\lambda_{m}^{\frac{3}{2}}
+\frac{5}{16}\lambda_{m}^{\frac{1}{2}}
+\frac{5}{16}
\qquad
(\lambda_{m}=1).
\end{eqnarray}
Substituting expressions (C-9), (C-12) and (C-13)
into Eq. (C-3) and using Eq. (45), we arrive
at Eqs. (49) to (51).

\section*{Appendix D}
\renewcommand{\theequation}{D-\arabic{equation}}
\setcounter{equation}{0}

It follows from Eq. (64) that for any $m=1,2,3$,
\begin{eqnarray*}
\ln P(Q^{2})-\ln P(Q^{2}
(1+(\lambda_{m}-1)z^{2}) &=&
\beta \Bigl (\frac{Q}{R}\Bigr )^{2\delta}
\Bigl [ \Bigl (1+(\lambda_{m}-1)z^{2}\Bigr )^{\delta}
-1\Bigr ]
\nonumber\\
&& -\alpha \ln (1+(\lambda_{m}-1)z^{2}),
\nonumber\\
\ln P(Q^{2})-\ln P(Q^{2}
(1+(\lambda_{m}^{-1}-1)z^{2}) &=&
\beta \Bigl (\frac{Q}{R}\Bigr )^{2\delta}
\Bigl [ \Bigl (1+(\lambda_{m}^{-1}-1)z^{2}\Bigr )^{\delta}
-1\Bigr ]
\nonumber\\
&& -\alpha \ln (1+(\lambda_{m}^{-1}-1)z^{2}).
\end{eqnarray*}
Substitution of these expressions into Eq. (47)
implies that
\begin{equation}
W^{(m)}=\Psi_{1}^{(m)}-\Psi_{2}^{(m)}
\end{equation}
with
\begin{eqnarray}
\Psi_{1}^{(m)} &=& k_{\rm B}T \biggl \{
-\frac{1}{2} (2a-1) \ln \lambda_{m}
+ K \biggl [ \int_{0}^{1} \biggl (a
(1+(\lambda_{m}-1)z^{2})^{\delta}
\nonumber\\
&& +(1-a)(1+(\lambda_{m}^{-1}-1)z^{2})^{\delta}
\biggr )dz-1\biggr ]\biggr \},
\nonumber\\
\Psi_{2}^{(m)} &=& k_{\rm B}T K_{1}\int_{0}^{1}
\Bigl [ a\ln (1+(\lambda_{m}-1)z^{2})
+(1-a)\ln (1+(\lambda_{m}^{-1}-1)z^{2})\Bigr ]dz,
\end{eqnarray}
where we set
\begin{equation}
K=\frac{4\pi\beta}{R^{2\delta}}
\int_{0}^{\infty} P(Q^{2})Q^{2(1+\delta)} dQ,
\qquad
K_{1}=4\pi\alpha \int_{0}^{\infty} P(Q^{2})Q^{2} dQ.
\end{equation}
Introducing the new variable $z=Q/R$ in the first
equality in Eq. (D-3) and using Eq. (7) to evaluate
the other equality, we find that
\begin{equation}
K=\beta\frac{\int_{0}^{\infty} \exp (
-\beta z^{3})z^{2\alpha+5}dz}
{\int_{0}^{\infty} \exp(-\beta z^{3})
z^{2\alpha +2}dz},
\qquad
K_{1}=\alpha.
\end{equation}
To determine the coefficient $K$, we set
$y=\beta^{\frac{1}{3}}z$, which results in
\[
K=\frac{\int_{0}^{\infty} \exp (-y^{3})y^{2\alpha+5}dy}
{\int_{0}^{\infty} \exp(-\beta y^{3})y^{2\alpha +2}dy}.
\]
Calculating the integral in the nominator by parts,
we arrive at the formula
\begin{equation}
K=1+\frac{2}{3}\alpha .
\end{equation}
The function $\Psi_{1}^{(m)}$ has already been found
in Appendix C,
\begin{eqnarray}
\Psi_{1}^{(m)} &=& k_{\rm B}T \biggl \{
-\frac{1}{2} (2a-1) \ln \lambda_{m}
+K\biggl [
a \Bigl (\frac{1}{6}\lambda_{m}^{\frac{5}{2}}
+\frac{5}{24}\lambda_{m}^{\frac{3}{2}}
+\frac{5}{16}\lambda_{m}^{\frac{1}{2}}\Bigr )
+(1-a)
\nonumber\\
&\times& \Bigl (\frac{1}{6}
\lambda_{m}^{-\frac{5}{2}}
+\frac{5}{24}\lambda_{m}^{-\frac{3}{2}}
+\frac{5}{16}\lambda_{m}^{-\frac{1}{2}}\Bigr )
+\frac{5}{16}\Bigl (aG(\lambda_{m})
+(1-a)G(\lambda_{m}^{-1})\Bigr )-1\biggl ]
\biggr \}.
\end{eqnarray}
To determine the function $\Psi_{2}^{(m)}$,
we introduce the notation
\begin{equation}
M(\mu)=\int_{0}^{1} \ln (1+\mu z^{2})dz
\end{equation}
and calculate the function $M(\mu)$ for
$\mu=\lambda_{m}-1$ and $\lambda_{m}^{-1}-1$,
respectively.
Performing integration by parts in Eq. (D-7),
we obtain
\begin{equation}
M(\mu)=z\Bigl [\ln (1+\mu z^{2})-2\Bigl ]_{z=0}^{z=1}
+2\int_{0}^{1} \frac{dz}{1+\mu z^{2}}
=\ln (1+\mu)+2\Bigl [
\int_{0}^{1} \frac{dz}{1+\mu z^{2}}-1\Bigr ].
\end{equation}
Calculating the integral, we find that
\[
\begin{array}{ll}
M(\mu)=\ln (1+\mu)+2\Bigl (
\frac{\arctan \sqrt{\mu}}{\sqrt{\mu}}-1\Bigr ),
& \mu>0,\\
M(\mu)=0, & \mu=0,\\
M(\mu)=\ln (1+\mu)+2\Bigl (
\frac{1}{2\sqrt{\mu}}\ln\frac{1+\sqrt{\mu}}{1
-\sqrt{\mu}}-1\Bigr ),
& \mu<0.
\end{array}
\]
This implies that
\begin{equation}
\int_{0}^{1} \ln (1+(\lambda_{m}-1)z^{2})dz=\ln \lambda_{m}
+2 H(\lambda_{m}),
\end{equation}
where the function $H(z)$ is given by Eq. (66).
Substitution of expression (D-9) into the second
equality in Eq. (D-2) results in
\begin{equation}
\Psi_{2}^{(m)}=k_{\rm B}TK_{1}\Bigl [ (2a-1)\ln \lambda_{m}
+2\Bigl (aH(\lambda_{m})+(1-a)H(\lambda_{m}^{-1})\Bigr )
\Bigr ].
\end{equation}
It follows from Eqs. (D-1), (D-4), (D-5), (D-6)
and (D-10) that
\begin{eqnarray}
W^{(m)}&=& k_{\rm B}T \biggl \{
-\frac{1}{2} (2a-1) \ln \lambda_{m}
+\Bigl (1+\frac{2}{3}\alpha\Bigr ) \biggl [
a \Bigl (\frac{1}{6}\lambda_{m}^{\frac{5}{2}}
+\frac{5}{24}\lambda_{m}^{\frac{3}{2}}
+\frac{5}{16}\lambda_{m}^{\frac{1}{2}}\Bigr )
+(1-a)
\nonumber\\
&\times& \Bigl (\frac{1}{6}
\lambda_{m}^{-\frac{5}{2}}
+\frac{5}{24}\lambda_{m}^{-\frac{3}{2}}
+\frac{5}{16}\lambda_{m}^{-\frac{1}{2}}\Bigr )
+\frac{5}{16}\Bigl (aG(\lambda_{m})
+(1-a)G(\lambda_{m}^{-1})\Bigr )-1\biggl ]
\biggr \}
\nonumber\\
&&-k_{\rm B}T \alpha \Bigl [ (2a-1)\ln \lambda_{m}
+2\Bigl (aH(\lambda_{m})+(1-a)H(\lambda_{m}^{-1})\Bigr )
\Bigr ].
\end{eqnarray}
Equation (65) follows from Eqs. (45) and (D-11).

\section*{Appendix E}
\renewcommand{\theequation}{E-\arabic{equation}}
\setcounter{equation}{0}

According to Eq. (52), the function $W_{0}$ is
given by
\begin{eqnarray}
W_{0} &=& k_{\rm B}T \biggl [
a \Bigl (\frac{1}{6}{\cal I}_{1}({\bf C}^{\frac{5}{2}})
+\frac{5}{24} {\cal I}_{1}({\bf C}^{\frac{3}{2}})
+\frac{5}{16}{\cal I}_{1}({\bf C}^{\frac{1}{2}})\Bigr )
\nonumber\\
&&+(1-a)
\Bigl (\frac{1}{6}{\cal I}_{1}({\bf C}^{-\frac{5}{2}})
+\frac{5}{24} {\cal I}_{1}({\bf C}^{-\frac{3}{2}})
+\frac{5}{16}{\cal I}_{1}({\bf C}^{-\frac{1}{2}})\Bigr )
-\frac{33}{16}\biggr ].
\end{eqnarray}
Expanding the function $H(z)$ in Eq. (66)
into a Taylor series in the vicinity of
the point $z=1$, we find that
\begin{eqnarray*}
H(z) = \sum_{k=1}^{\infty} \frac{(-1)^{k} (z-1)^{k}}{2k+1}
\quad (z>1),
\qquad
H(z) = \sum_{k=1}^{\infty} \frac{(1-z)^{k}}{2k+1}
\quad (z<1),
\end{eqnarray*}
which means that for any $z$ belonging to the region
where the series converges,
\begin{equation}
H(z) = \sum_{k=1}^{\infty} \frac{(1-z)^{k}}{2k+1}.
\end{equation}
It follows from Eq. (E-2) that
\[
\sum_{m=1}^{3} H(\lambda_{m})
=\sum_{k=1}^{\infty} \frac{1}{2k+1}
\sum_{m=1}^{3} (1-\lambda_{m})^{k}
=\sum_{k=1}^{\infty} \frac{1}{2k+1}
{\cal I}_{1}(({\bf I}-{\bf C})^{k}).
\]
Similarly,
\[
\sum_{m=1}^{3} H(\lambda_{m}^{-1})
=\sum_{k=1}^{\infty} \frac{1}{2k+1}
{\cal I}_{1}(({\bf I}-{\bf C}^{-1})^{k}).
\]
Substitution of these expressions into Eq. (67)
results in
\begin{eqnarray}
W &=&  \Bigl (1+\frac{2}{3}\alpha\Bigr )W_{0}
- k_{\rm B}T \Bigl [ (2a-1) (\alpha+\frac{1}{2})\ln J_{3}
\nonumber\\
&& +2\alpha \sum_{k=1}^{\infty} \frac{1}{2k+1}
\Bigl (a {\cal I}_{1}(({\bf I}-{\bf C})^{k})
+(1-a){\cal I}_{1}(({\bf I}-{\bf C}^{-1})^{k})
\Bigr ) \Bigr ].
\end{eqnarray}
Equation (68) follows from Eqs. (E-1) and Eq. (E-3)
and definition (69) of the strain tensors.

As the series in Eq. (E-3) converges rather slowly,
it is tempting to truncate it up to terms of
an order $k_{0}$ with respect to the norms
of the strain tensors.
Taking into account one term only ($k_{0}=1$),
we find that
\begin{eqnarray}
&& \sum_{k=1}^{k_{0}} \frac{1}{2k+1}
\Bigl [ a {\cal I}_{1}(({\bf I}-{\bf C})^{k})
+(1-a){\cal I}_{1}(({\bf I}-{\bf C}^{-1})^{k}))
\Bigr ]
\nonumber\\
&& =-\frac{1}{3}\Bigl [ a \Bigl ({\cal I}_{1}({\bf C})-3\Bigr )
+(1-a)\Bigl ( {\cal I}_{1}({\bf C}^{-1})-3\Bigr )\Bigr ].
\end{eqnarray}
It follows from Eq. (E-4) that in the first-order
approximation, the account for the pre-exponential
term in the distribution function (6) is
equivalent to the introduction of
additional Gaussian chains into the network.

Taking into account two terms ($k_{0}=2$) in
Eq. (E-3), we find, after simple algebra, that
\begin{eqnarray}
&& \sum_{k=1}^{k_{0}} \frac{1}{2k+1}
\Bigl [ a {\cal I}_{1}(({\bf I}-{\bf C})^{k})
+(1-a){\cal I}_{1}(({\bf I}-{\bf C}^{-1})^{k}))
\Bigr ]
\nonumber\\
&&= \frac{1}{15}\biggl \{a \Bigl [ 3(J_{1}-3)^{2}
+7 (J_{1}-3)-6 (J_{2}-3)\Bigr ]
\nonumber\\
&& +(1-a)\Bigl [ 3(J_{-1}-3)^{2}
+7 (J_{-1}-3)-6 (J_{-2}-3)\Bigr ]\biggr \}.
\end{eqnarray}
According to Eq. (E-5), in the second-order
approximation, the account for the pre-exponential
term in Eq. (6) results in a decrease in the
strain energy of a self-avoiding chain.

\newpage

\section*{List of figures}
\parindent 0 mm

{\bf Figure 1:}
The dimensionless strain energy
$\overline{W}$ versus elongation ratio $k$
at uniaxial extension of an incompressible network.
Solid line: the Ogden model (52).
Circles: the Ogden model with account for
the correction term (50).
\vspace*{2 mm}

{\bf Figure 2:}
The dimensionless strain energy
$\overline{W}$ versus shear $k$ at simple shear
of an incompressible network.
Solid line: the Ogden model (52).
Circles: the Ogden model with account for
the correction term (50).
\vspace*{2 mm}

{\bf Figure 3:}
The engineering stress $\sigma_{\rm e}$
versus elongation ratio $k$ at uniaxial
extension of natural rubber with the concentration
of cross-linker $\phi=1.0$, 1.5, 2.0, 2.5, 3.0,
3.5 and 4.0 phr, from bottom to top,
respectively.
Circles: experimental data \cite{KS00}.
Solid lines: results of numerical simulation.
\vspace*{2 mm}

{\bf Figure 4:}
The elastic modulus $\mu_{1}$
versus concentration of cross-linker $\phi$.
Circles: treatment of observations \cite{KS00}.
Solid line: approximation of the experimental data
by Eq. (63) with $\mu_{1}^{(1)}=0.12$ MPa/phr.
\vspace*{2 mm}

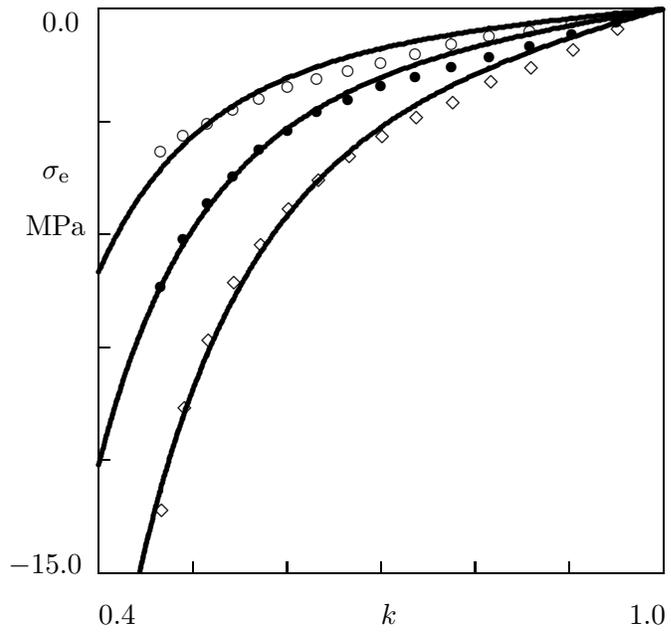
{\bf Figure 5:}
The engineering stress $\sigma_{\rm e}$
versus elongation ratio $k$ at uniaxial compression
of chloroprene rubber reinforced with carbon black (CB).
Symbols: experimental data \cite{BB98}.
Unfilled circles: 15 phr CB.
Filled circles: 40 phr CB.
Diamonds: 65 phr CB.
Solid lines: results of numerical simulation.
\vspace*{2 mm}

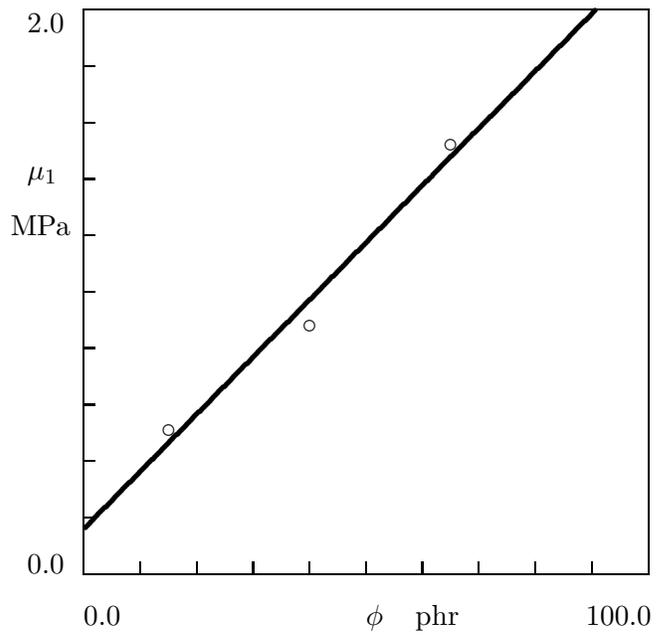
{\bf Figure 6:}
The elastic modulus $\mu_{1}$
versus concentration of carbon black $\phi$
for chloroprene rubber.
Circles: treatment of observations \cite{BB98}.
Solid line: approximation of the experimental data
by Eq. (63) with $\mu_{1}^{(0)}=0.16$ MPa and
$\mu_{1}^{(1)}=0.02$ MPa/phr.
\vspace*{2 mm}

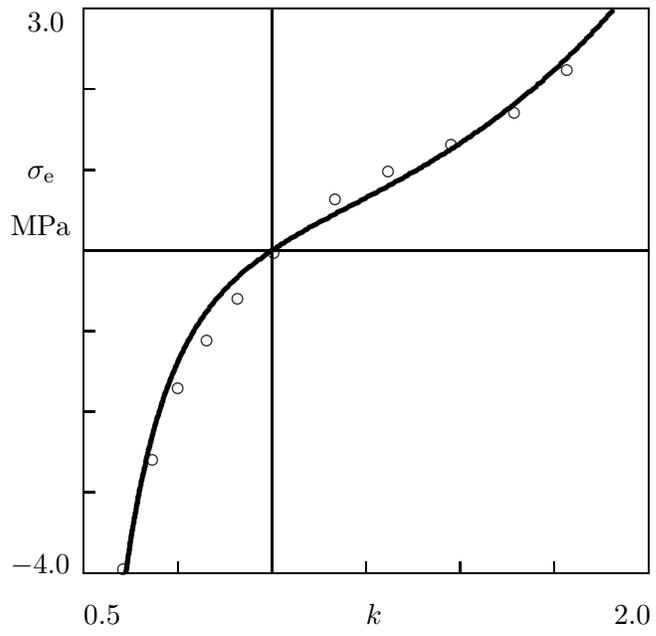
{\bf Figure 7:}
The engineering stress $\sigma_{\rm e}$
versus elongation ratio $k$ at uniaxial tension--compression
of carbon black-filled natural rubber.
Circles: experimental data \cite{HTS03}.
Solid line: results of numerical simulation
with $\mu_{1}=0.39$ and $\mu_{2}=0.27$ MPa.
\vspace*{2 mm}

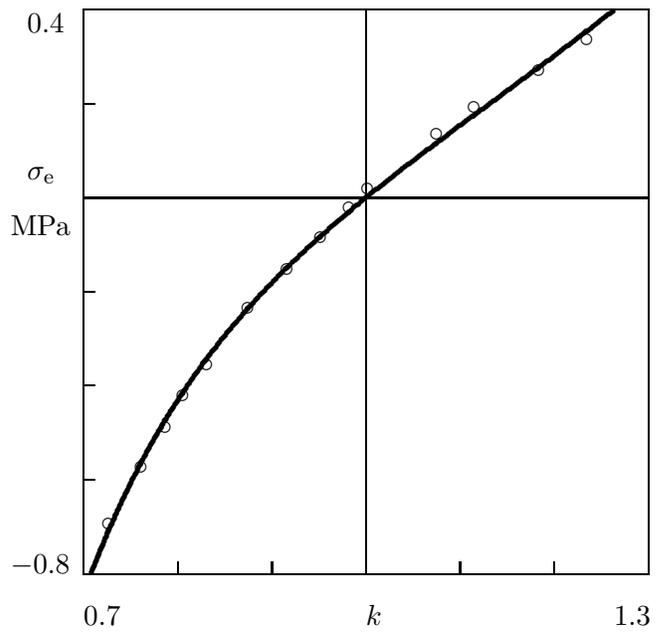
{\bf Figure 8:}
The engineering stress $\sigma_{\rm e}$
versus elongation ratio $k$ at uniaxial tension--compression
of polybutadiene rubber.
Circles: experimental data \cite{RMH99}.
Solid line: results of numerical simulation
with $\mu_{1}=0.285$ and $\mu_{2}=0.056$ MPa.
\vspace*{2 mm}

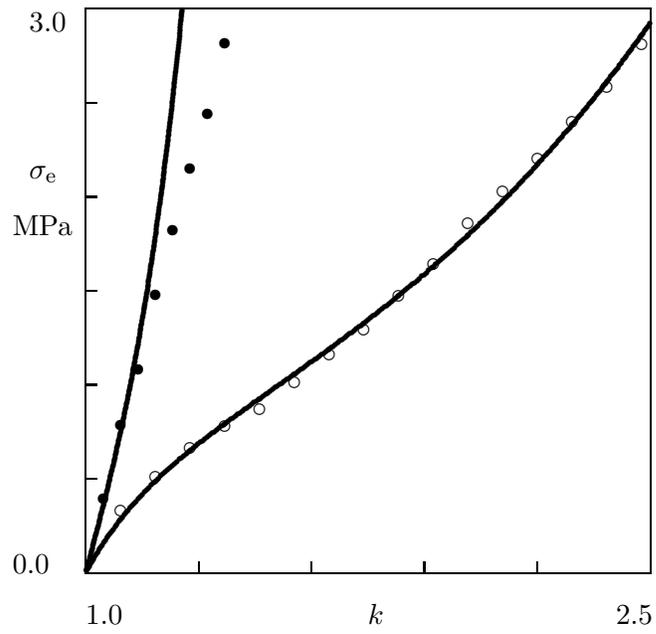
{\bf Figure 9:}
The engineering stress $\sigma_{\rm e}$
versus elongation ratio $k$ at uniaxial
(unfilled circles) and equi-biaxial (filled circles)
extension of synthetic rubber Smactane.
Symbols: experimental data \cite{CCH01}.
Solid lines: results of numerical simulation
with $\mu_{1}=0.075$ and $\mu_{2}=0.704$ MPa.
\vspace*{2 mm}

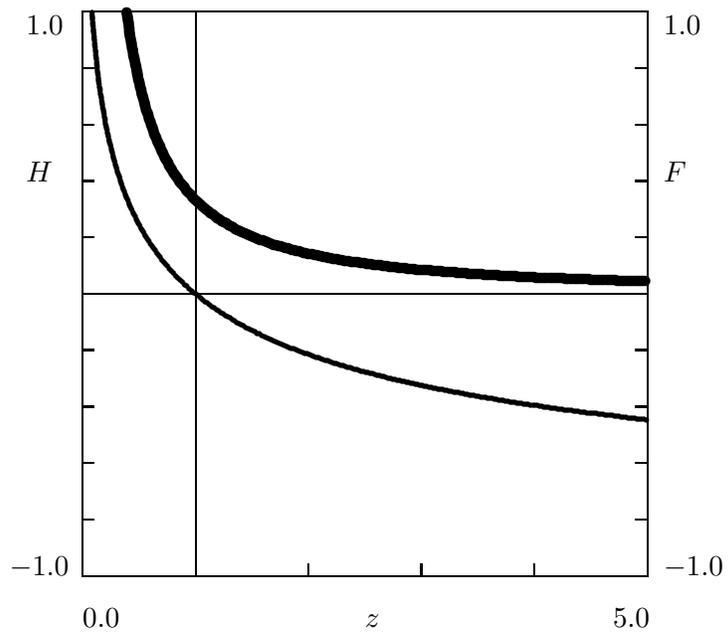
{\bf Figure 10:}
The functions $H(z)$ (thin curve) and $F(z)$ (thick curve).
\vspace*{2 mm}

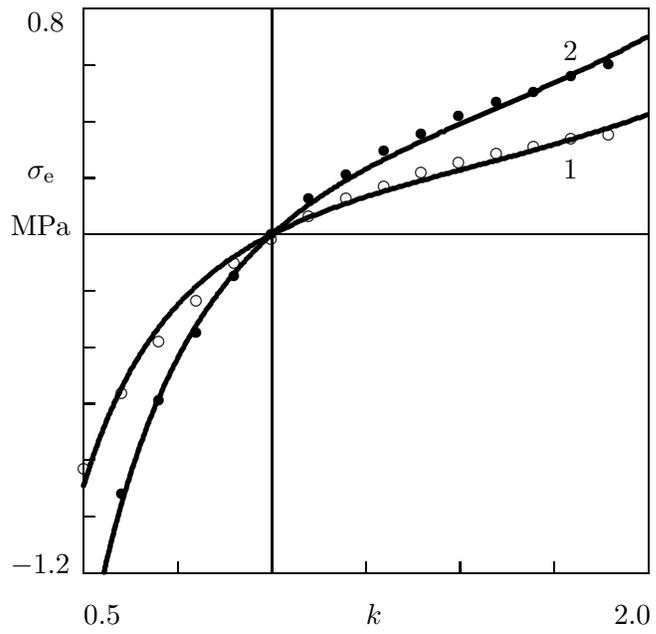
{\bf Figure 11:}
The engineering stress $\sigma_{\rm e}$
versus elongation ratio $k$ at uniaxial tension--compression
of carbon black-filled natural rubbers
NR--1 (unfilled circles) and NR--2 (filled circles).
Symbols: experimental data \cite{RMH99}.
Solid lines: results of numerical simulation
with $a=1.0$.
Curve 1: $\mu_{1}=0.0296$, $\mu_{2}=-2.72$ MPa.
Curve 2: $\mu_{1}=0.0449$, $\mu_{2}=-4.94$ MPa.
\vspace*{2 mm}

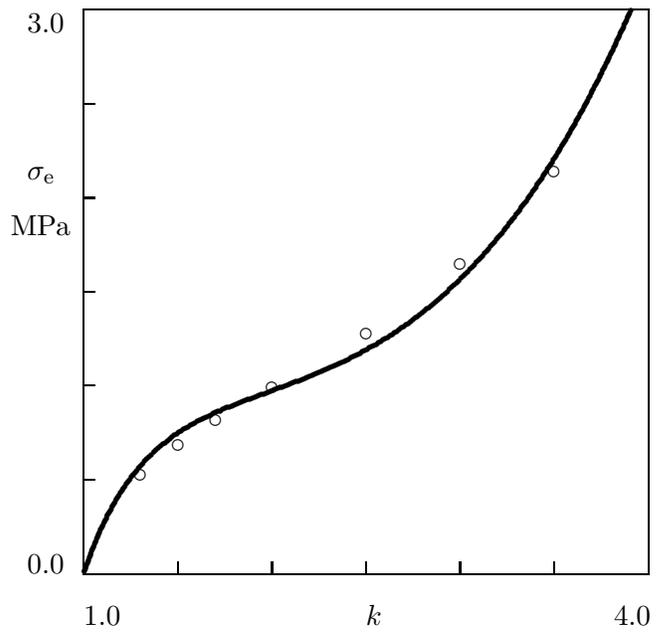
{\bf Figure 12:}
The engineering stress $\sigma_{\rm e}$
versus elongation ratio $k$ at equi-biaxial
extension of natural rubber.
Circles: experimental data \cite{JGS75}.
Solid line: results of numerical simulation
with $a=1.0$, $\mu_{1}=0.023$ and $\mu_{2}=-6.96$ MPa.
\vspace*{2 mm}

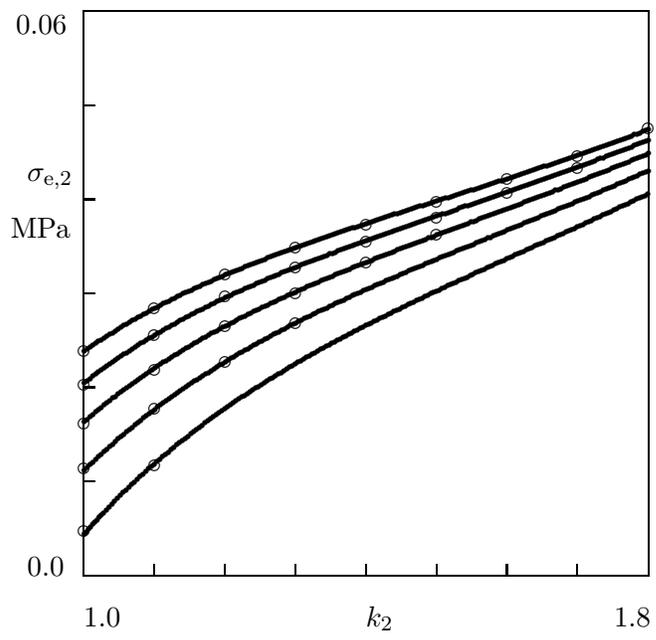
{\bf Figure 13:}
The engineering stress $\sigma_{{\rm e},2}$
versus elongation ratio $k_{2}$ at biaxial tension
of PDMS network with the elongation ratios
$k_{1}=1.1$, 1.3, 1.5, 1.7 and 1.9, from bottom
to top, respectively.
Circles: experimental data \cite{KUK02}.
Solid lines: results of numerical simulation
with $a=1.0$, $\mu_{1}=0.0027$ and $\mu_{2}=-0.291$ MPa.

\setlength{\unitlength}{0.75 mm}
\begin{figure}[tbh]
\begin{center}

\end{center}
\vspace*{10 mm}

\caption{}
\end{figure}

\setlength{\unitlength}{0.75 mm}
\begin{figure}[tbh]
\begin{center}
%
\end{center}
\vspace*{10 mm}

\caption{}
\end{figure}

\setlength{\unitlength}{0.75 mm}
\begin{figure}[tbh]
\begin{center}
%
\end{center}
\vspace*{10 mm}

\caption{}
\end{figure}

\setlength{\unitlength}{0.75 mm}
\begin{figure}[tbh]
\begin{center}
%
\end{center}
\vspace*{10 mm}

\caption{}
\end{figure}
\end{document}